\documentclass[11pt]{article}
\usepackage{mathrsfs}
\usepackage{amsfonts}
\usepackage{amsmath,amsthm,hyperref}
\usepackage{amssymb}
\usepackage{hyperref,color}
\usepackage{exscale,relsize}
\usepackage[all]{xy}
\usepackage{exscale,relsize,graphicx,caption}

\makeatletter
\def\equfill@{\arrowfill@\Relbar\Relbar\Relbar}
\newcommand{\equfill}[2][]{\ext@arrow 0395\equfill@{#1}{#2}}
\makeatother
\makeatletter
\def\Eqlfill@{\arrowfill@\Relbar\Relbar\Relbar}
\newcommand{\extendEql}[1][]{\ext@arrow 0359\Eqlfill@{#1}}
\makeatother

\newcommand{\beq}{\begin{equation}}
\newcommand{\eeq}{\end{equation}}
\newcommand{\baq}{\begin{eqnarray}}
\newcommand{\eaq}{\end{eqnarray}}

\numberwithin{equation}{section}
\begin{document}

\author{ Peng Zhao \footnote{College of Art and Sciences, Shanghai Maritime
University, Shanghai 201306, P.~R.~China}   \quad \quad Engui Fan \footnote{School of Mathematical Sciences, Institute of Mathematics
 and Key Laboratory of Mathematics for Nonlinear Science,  Fudan
University, Shanghai 200433, P.~R.~China}~\footnote{Corresponding
author and  e-mail address:
      faneg@fudan.edu.cn} \and }
\date{}
\title{\bf \Large{ New Construction of Algebro-Geometric Solutions to
the Modified Kadomtsev-Petviashvili Hierarchy} }
\maketitle
\begin{abstract}
We extend  Gesztesy-Holden's method  to 2+1 dimensional case  to  obtain  a unified construction
to the algebro-geometric solutions of the whole modified Kadomtsev-Petviashvili (mKP) hierarchy.
Our tools include the relations between solutions of the Gerdjikov-Ivanov (GI) and mKP hierarchy,
the Baker-Akhiezer functions in 2+1 dimensions, a special function
 $\psi_1(P)\psi_2(P^*)$ on $X\times \mathbb{R}^3$ and
 Dubrovin-type equations for auxiliary divisors.

\end{abstract}

\section{Introduction}
The KP hierarchy and its counterpart, for example, the mKP hierarchy,
plays an important role in a variety of
different fields including modern string theory and in connection with the solution of the
Schottky problem of compact Riemann surfaces \cite{Mul1988},~\cite{segal1985}.
It is generally believed that all integrable hierarchies of (1+1)-dimensional equations 
by means of the inverse scattering method can be represented as certain reductions of a universal
Kadomtsev-Petviashvili (KP) hierarchy of (2+1)-dimensional equations and/or of its extentions
to modified and multicomponent cases \cite{Hirota1986},~\cite{Sach1988}.
 This hypothesis originates from a unifying Sato theory
that describes the KP hierarchy in terms of the pseudodifferential operator 
$$\mathscr{L}=\frac{\partial}{\partial x}+ \sum_{j=1}^{\infty} u_{j+1}(\frac{\partial}{\partial x})^{-j}$$
 with functions
$u_j=u_j(t_1,t_2,\ldots)$ that depends on an
infinite number of independent variables $t_j,j\in\mathbb{N}$ \cite{dmmt1981},~\cite{dkjm1983},~\cite{JM1983},~\cite{sato},
which indicates that by introducing an infinite set of ¡®time¡¯ variables one can also treat the integrable
equations as flows on infinite-dimensional Grassmannian manifolds \cite{JM1983}.
Based on the treatment of partial differential KP equations as dynamical systems on the infinite-dimensional algebra of pseudo-differential operators \cite{a1},
the KP equation \cite{dkjm1983},~\cite{JM1983}
\begin{equation}\label{1.1kp}
  u_t=\frac{1}{4}u_{xxx}+3uu_x+\frac{3}{4}\partial^{-1}_xu_{yy},
\end{equation}
 is considered as the simplest member of a hierarchy of
equations that can be brought to bilinear form and solved by the $\tau$-funtion approach.
 This
theory reveals deep interrelations between the Hamiltonian structures of the KP hierarchy
and two-dimensional conformal field theory as well as W$_{1+\infty}$ algebras~\cite{Yamagishi}.

The problem of constructing the quasiperiodic solutions (QPS) is one of the most challenging
problems of the theory of integrable systems, and many mathematicians and physicists
spent much efforts to obtain the QPS for almost all equations that are known to be integrable.
Among the most powerful solution generating methods are techniques
from algebraic geometry which lead to solutions in terms of Riemann theta functions on
certain Riemann surfaces, that is, the so-called algebro-geometric solutions.
This kinds of solutions were originally studied on the KdV equation based on
the inverse spectral theory and algebro-geometric method developed by pioneers such as Novikov~\cite{DubNov}, Dubrovin \emph{et al.}~\cite{dmn1976},~\cite{Novikov1976m}, Its \& Matveev~~\cite{itsma1975},~\cite{Its1975}, Lax~\cite{lax1975},
and McKean \& van Moerbeke~\cite{mckean1975} for 1 + 1 systems, and
extended by Krichever in 1976 for 2 + 1 systems like KP~\cite{Krichever1976},~\cite{Krichever1977a} in the late 1970s.
Later this theory has
been developed to the whole hierarchies of nonlinear integrable equations by Gesztesy,
Holden \emph{et al.} using polynomial recursion method \cite{gesztesy1999},~\cite{gesztesy1998},~\cite{gesztesy2003a},~\cite{Gesztesyhoden2003}.
 Another breakthrough in this area was made by Mumford~\cite{Mumford} in early 1980s, who observed  that integrable equations like KdV, KP or sine-Gordon, are hidden in Fay's trisecant formula. Mumford's approach are based on degenerated versions
of Fay's identity, which reveals the relations between algebro-geometric solutions of integrable equations and a purely algebro-geometric identity  \cite{ckalla2013}, \cite{clein2002}, \cite{clein2005}.
A detailed introduction and recent development about this subject can be found in the survey article \cite{Matveev2008}.


The study of higher dimensional integrable equations is usually considered a more difficult problem than (1+1)-dimensional equations and
by now, much work has been done on KP and mKP equation.
An important development of the algebro-geometric method
was the passage from 1+1 systems to the integration of 2+1 KP-like systems, realized by Krichever
in 1976 \cite{Krichever1976}, who constructed algebro-geometric solutions of
the KP equation on a
basis of a purely algebraic formulation of algebro-geometric approach.
A generalized Miura transformation between the KP equation and mKP equation
was discussed in \cite{gest1991}.
Later it has been shown that
the mKP hierarchy
can be constructed on the basis of the Sato approach by means of the gauge
transformation of the KP hierarchy which results in modification of the pseudo-differential operator
$\mathscr{L}$ \cite{Oevel1993a}. In ref. \cite{gest1998}, the authors obtain new B\"{a}cklund transformations for the
KP hierarchy and the possibility of transferring classes of KP solutions into those of
mKP solutions. A explicit theta function solution of the mKP equation is derived by the technique of
nonlinearization of Lax pairs and Abel inversion \cite{chengeng}. However, within the knowledge of
the authors, the algebro-geometric solutions of the entire mKP hierarchy has not been considered so far.

In this paper, using inherent relations between the mKP hierarchy and GI hierarchy, we
shall improve the method in \cite{Gesztesyhoden2003} to obtain  a unified construction
to the algebro-geometric solutions of the whole hierarchy.
We extend the concepts the Baker-Ahkiezer function to the mKP hierarchy and find
a class of quasiperiodic solutions by algebro-geometric method.
The construction exploits explicit form of the Baker-Ahkiezer function
and involve also a fundamental meromorphic function, which
allows for the analytic properties of the Baker-Ahkiezer function.
The theta function representation for algebro-geometric solutions of the mKP hierarchy are then verified directly using formulae for the Baker-Ahkiezer function
and its $x$-derivative.
In our construction,
theoretical analysis of algebro-geometric solution for the whole mKP hierarchy are carried out from a different angle of view and
the whole approach discussed in the present paper is a general one and gives
similar results for other 2+1 dimensional and higher dimensional soliton equations.

This paper is organized as follows. In section 2,
we first introduce the GI hierarchy and mKP hierarchy and then discuss the Burchnall-Chaundy polynomial
in connection with a basic initial value problem and underlying hyperelliptic curve.
In section 3,
the dynamics of auxiliary spectral points $\{\mu_j\}_{j=1}^n, \{\nu_j\}_{j=1}^n$ and corresponding trace formula
is considered.
In section 4, we present explicit
representation for the
Baker-Akhiezer functions $\psi_1(P), \psi_2(P)$, whose
analytic properties can be derived by studying an important function $\psi_1(P)\psi_2(P^*)$.
In section 5, we shall obtain Riemann
theta function representation for $\psi_1(P),\psi_2(P), \psi_1(P)\psi_2(P^*)$,
and especially for the algebro-geometric solutions $q,r$ of the whole mKP hierarchy.

\section{mKP Hierarchy, GI Hierarchy, Burchnall-Chaundy Polynomial and Basic Initial Value Problem}

To make this paper self-contained,
we shall first introduce the mKP hierarchy in standard literature
and then provide the construction of the GI hierarchy and derive the associated sequence of Lax pairs using a polynomial recursion formalism.
Moreover, we obtain the Burchnall-Chaundy polynomial in connection with the modified KP
hierarchy and underlying hyperelliptic curve.

Throughout this section we make the following hypothesis.

\newtheorem{hyp1}{Hypothesis}[section]
\begin{hyp1}
Suppose that $q,r: \mathbb{R}^3 \rightarrow \mathbb{C}$
satisfy
\begin{equation}\label{hp1}
\begin{split}
  &q(\cdot,y,t_{p})\in C^\infty(\mathbb{R}),~~t_{p}\in\mathbb{R},\quad \\ &q(x,y,\cdot),r(x,y,\cdot),q(x,\cdot,t_p),r(x,\cdot,t_p)\in C^1(\mathbb{R}),~~x,y\in\mathbb{R},\\
  &q(x,y,t_{p})\neq 0,~~r(x,y,t_{p})\neq 0,~~(x,y,t_{p})\in\mathbb{R}^3.
  \end{split}
\end{equation}

\end{hyp1}

The KP and mKP equation, e.g.
\cite{JM1983},~\cite{Kash1981},~\cite{kon1982},~\cite{kon1984}) and the algebraic framework of the KP and mKP hierarchy can be found in \cite{a1},~\cite{al1981},~\cite{Dickey1991},~\cite{gest1994},
~\cite{Mul1984},~\cite{Pre1989},~\cite{Kuperschmidt1981},~\cite{Wilson1985}.
In Sato theory, the mKP hierarchy is described by the
isospectral deformations of the eigenvalue problem
\begin{equation}\label{st2.1}
\begin{split}
 & L\psi=\lambda \psi,~~ \lambda\in\mathbb{C},
  \end{split}
\end{equation}
where the pseudodifferential operator $L$ is given by
\begin{equation}\label{sato1}
L= \partial+ \sum_{j=0}^{\infty} u_{j+1}\partial^{-j},~~ \partial=\frac{\partial}{\partial x},
\end{equation}
and $u_j$ are functions in
infinitely many variables $(x,t_1,t_2\ldots)$ with $t_1=y.$
We denote by $B_m$ the differential part of $L^m:$
\begin{align}\label{sato2}
   B_m= (L^m)_+=\sum_{j=1}^m b_{m,j}\partial^j.
\end{align}
The coefficients $b_{m,j}$ in (\ref{sato2}) can be uniquely determined
by the coordinates $u_j$, and their $x$ derivatives.
Explicitly,
\begin{align}
  B_1=&~\partial, \nonumber\\
  B_2=&~\partial^2+2u_1 \partial,\nonumber\\
  B_3=&~\partial^3+3u_1\partial^2+3 (u_2+u_{1,x}+u_{1}^2)\partial,\nonumber\\
  B_4=&~\partial^4+4u_1\partial^3+(4 u_2+6u_{1,x}+6u_1^2)\partial^2+(4u_3+6u_{2,x}\nonumber\\
   &+4u_{1,xx}+12u_1u_2+12u_1u_{1,x}+4u_1^3)\partial,~~\textrm{etc.} \nonumber
\end{align}
From the compatibility conditions of (\ref{st2.1}) and
\begin{equation}
  \phi_{t_m}=B_m\phi,
\end{equation}
we have
\begin{align}\label{st2.5}
  L_{t_m}=[B_m,L],
\end{align}
or equivalently,
\begin{equation}\label{st2.6}
  (B_{m})_{t_n}-(B_{n})_{t_m}=[B_n,B_m].
\end{equation}
The mKP hierarchy is obtained from (\ref{st2.5}) or (\ref{st2.6}) for
infinite coordinates $\{u_j\}_{j=2}^\infty.$ For example,
from (\ref{st2.5}), one obtains
\begin{align}
u_{1,t_2}=&~2u_{2,x}+u_{1,xx}+2u_1u_{1,x},\label{st2.7}\\
u_{2,t_2}=&~2u_{3,x}+u_{2,xx}+2(u_1u_{2})_x,\label{st2.8}\\
u_{3,t_2}=&~2u_{4,x}+u_{3,xx}+2u_{1}u_{3,x}+4u_{1,x}u_{3}-2u_{1,xx}u_2,\label{st2.9}\\
&\ldots\ldots\nonumber\\
u_{1,t_3}=&~3u_{3,x}+3u_{2,xx}+u_{1,xxx}+6(u_1u_{2})_x+3(u_1u_{1,x})_x+3u_1^2u_{1,x},\label{st2.10}\\
u_{2,t_3}=&~3u_{4,x}+3u_{3,xx}+u_{2,xxx}+6(u_1u_{3})_x+3(u_1u_{2,x})_x+6u_2u_{2,x}\nonumber\\
&~+3(u_1^2u_2)_x,\\
&\ldots\ldots\nonumber\\
u_{1,t_4}=&~4u_{4,x}+6u_{3,xx}+4u_{2,xxx}+u_{1,xxxx}+12(u_1u_3)_x+12u_2u_{2,x}\nonumber\\
&~+12(u_1u_{2,x})_x+6(u_{1,x}u_2)_x+4(u_1u_{1,xx})_x+6u_{1,x}u_{1,xx}\nonumber\\
&~+12(u_1^2u_2)_x+6(u_1^2u_{1,x})_x+4u_1^3u_{1,x} \label{st2.12}\\
&\ldots\ldots\nonumber
\end{align}
Eliminating $u_2$, $u_3$, $u_4$ from (\ref{st2.7}), (\ref{st2.8}), (\ref{st2.9}) and
taking into account
(\ref{st2.10}), (\ref{st2.12}), one obtains
   \begin{align}
      u_{t_3}=&~\frac{1}{4}u_{xxx}-\frac{3}{2}u^2u_x+\frac{3}{2}u_x\partial^{-1}u_y
      +\frac{3}{4}\partial^{-1}u_{yy}\label{stkp}\\
      u_{t_4}=&~\frac{1}{2}u_{xxy}-2u^2u_y+u\partial^{-1}u_{yy}+
      2u_y\partial^{-1}u_y-\partial^{-1}(uu_y)_y-2u_x\partial^{-1}uu_y\nonumber\\
      &~+u_x\partial^{-2}u_{yy}+\frac{1}{2}\partial^{-2}u_{yyy}.\label{stkp2}
\end{align}
where we denote by $u=u_1.$ Equation (\ref{stkp}) is
the mKP equation and equation (\ref{stkp2}) is the first higher order flows of the mKP hierarchy.
Similarly, the mKP and higher-order mKP equation can also be derived from (\ref{st2.6}) with
$n=2,m=3$ and $n=2, m>3$, respectively.

Next we construct the GI hierarchy. To this end one has to consider the following one-dimensional $2\times 2$
matrix-valued differential expression
\begin{equation}
  M=\left(
      \begin{array}{cc}
        -\partial+\frac{1}{4}qr & -qz \\
        rz &  \partial-\frac{1}{4}qr\\
      \end{array}
    \right),
\end{equation}
and another $2\times 2$ matrix-valued differential expression
of order $2n+2$ with respect to $z,$ denoted by $Q_{2n+2},n\in\mathbb{N}_0,$ which is defined recursively
in the following.
  Defines the sequences of differential polynomials $\{f_\ell\}_{\ell\in\mathbb{N}_0}, \{g_\ell\}_{\ell\in\mathbb{N}_0},$ and $\{h_\ell\}_{\ell\in\mathbb{N}_0},$ recursively by
 \begin{align}
  &g_{2\ell+1}=~f_{2\ell}=h_{2\ell}=0,~~\ell\in\mathbb{N}_0,\label{st2.13}\\
  &f_1=~2q,~g_0=2,~h_1=-2r,\\
  &2f_{2\ell+1}=-f_{2\ell-1,x}+2qg_{2\ell}+(1/2)qr f_{2\ell-1}, ~\ell\in\mathbb{N}_0,\\
& 2h_{2\ell+1}=~h_{2\ell-1,x}-2rg_{2\ell}+(1/2)qr h_{2\ell-1},~\ell\in\mathbb{N}_0,\\
&g_{2\ell,x}=~rf_{2\ell+1}+qh_{2\ell+1},~\ell\in\mathbb{N}_0.\label{st2.16}
 \end{align}
Here $q,r$ should be considered as functions of infinitely many variables $(t_{1}, t_{2},t_{3},\ldots)$.
Explicitly, one computes
\begin{align*}
  f_1=&~2q, \\
  f_3=&~-(1/2)q^2r-q_x+c_1(2q),\\
  f_5=&~(1/8)q^3r^2+(3/4)qq_xr-(1/4)q^2r_x+(1/2)q_{xx}\nonumber\\
  &~+c_1(-q^2r/2-q_x)+c_2(2q), \\
  &\ldots\ldots,\\
  g_0=&~2,\\
  g_2=&~-qr+2c_1,\\
  g_4=&~(1/4)q^2r^2+(1/2)(q_xr-qr_x)+c_1(-qr)+2c_2,\\
  g_6=&~-q^3r^3+3q^2rr_x-3qq_xr^2-q_{xx}r-qr_{xx}+q_xr_x \nonumber\\
  &+((1/4)q^2r^2+(1/2)(q_xr-qr_x))c_1+c_2(-qr)+2c_3, \\
  &\ldots\ldots,\\
  h_1=&~-2r,\\
  h_3=&~(1/2)qr^2-r_x+c_1(-2r),\\
  h_5=&~-(1/8)q^2r^3+(3/4)qrr_x-(1/4)q_xr^2-(1/2)r_{xx}\nonumber\\
  &+c_1(qr^2/2-r_x)+c_2(-2r),~~\textrm{etc.,}
\end{align*}
where $\{c_{\ell}\}_{\ell=0}^\infty\subset\mathbb{C}$ are integration constants. Subsequently,
we also introduce the corresponding homogeneous
coefficients $\hat{f}_\ell,\hat{g}_{\ell},$ and $\hat{h}_\ell$, defined by
vanishing of the integration constants $c_k,$ for $k=1,\ldots,\ell,$
\begin{align}
 &\hat{f}_1=2q, ~~\hat{f}_{2\ell+1}=f_{2\ell+1}|_{c_k=0,k=1,\ldots,\ell},~~\ell\in\mathbb{N},\\
 &\hat{g}_0=2,~~\hat{g}_{2\ell}=g_{2\ell}|_{c_k=0,k=1,\ldots,\ell},~~\ell\in\mathbb{N},\\
 &\hat{h}_1=-2r,~~\hat{h}_{2\ell+1}=h_{2\ell+1}|_{c_k=0,k=1,\ldots,\ell},~~\ell\in\mathbb{N}.
\end{align}
Hence,
\begin{equation}
\begin{split}
  &f_{2\ell+1}=\sum_{k=0}^\ell c_{\ell-k}\hat{f}_{2k+1},~~ g_{2\ell}=\sum_{k=0}^\ell c_{\ell-k}\hat{g}_{2k},\\
  &h_{2\ell+1}=\sum_{k=0}^\ell c_{\ell-k}\hat{h}_{2k+1},
  \end{split}
\end{equation}
introducing $c_0=1.$
Next we construct the stationary GI hierarchy. Now $q,r$ are considered as functions
of the variable $x$
and the $2\times 2$ matrix-valued differential expression $Q_{2n+2}$ is introduced by
\begin{equation}\label{st2.27}
  Q_{2n+2}=\sum_{j=0}^n\left(
             \begin{array}{cc}
               -g_{2\ell}z^2 & f_{2\ell+1}z \\
               -h_{2\ell+1}z & g_{2\ell+2}z^2 \\
             \end{array}
           \right)M^{n-\ell}
           +\left(
              \begin{array}{cc}
                \frac{1}{2}g_{2n+2} & 0 \\
                0 & -\frac{1}{2}g_{2n+2} \\
              \end{array}
            \right),~n\in\mathbb{N}_0,
\end{equation}
Using the recursion relations (\ref{st2.13})-(\ref{st2.16}), we rewrite the commutator of $Q_{2n+2}$
and $M$ as
\beq\label{st2.28}
\begin{split}
[Q_{2n+2},M]=&~[Q_{2n+2},M]|_{\textrm{ker}(M-z^2)}\\
 =&\left(
        \begin{array}{cc}
          -\frac{1}{2}g_{2n+2,x} & -2f_{2n+3}+q g_{2n+2} \\
          -2h_{2n+3}-r g_{2n+2} & \frac{1}{2}g_{2n+2,x} \\
        \end{array}
      \right),~n\in
      \mathbb{N}_0,
 \end{split}
\eeq
where
\beq\label{st2.28d}
\textrm{ker}(M-z^2)=\{\Psi=\left(
                                \begin{array}{c}
                                  \psi_1 \\
                                  \psi_2 \\
                                \end{array}
                              \right):~\mathbb{R}\rightarrow \mathbb{C}_{\infty}^2|(M-z^2)\Psi=0
\},~~z\in\mathbb{C},
\eeq
denotes the two-dimensional kernel of $M-z^2$.
The stationary GI hierarchy can be constructed
in terms of the vanishing of the commutator of $Q_{2n+2}$
and $M$ in
(\ref{st2.28}), that is,
\beq\label{st2.28c}
[Q_{2n+2},M]=0,~~n\in\mathbb{N}_0,
\eeq
or equivalently,
\beq\label{f2.29}
\left(
        \begin{array}{cc}
          -\frac{1}{2}g_{2n+2,x} & -2f_{2n+3}+q g_{2n+2} \\
          -2h_{2n+3}-r g_{2n+2} & \frac{1}{2}g_{2n+2,x} \\
        \end{array}
      \right)=0,~~n\in
      \mathbb{N}_0.
\eeq
From $2f_{2n+3}-q g_{2n+2}=0$ and $2h_{2n+3}+r g_{2n+2}=0,$
we get
\beq
2rf_{2n+3}-qr g_{2n+2}=0,~~2qh_{2n+3}+qr g_{2n+2}=0,
\eeq
that is,
\beq
2(rf_{2n+3}+qh_{2n+3})=0,
\eeq
which implies $\frac{1}{2}g_{2n+2,x}=0.$
Thus, varying $n\in\mathbb{N}_0,$ stationary GI hierarchy (\ref{f2.29}) can be written as
\begin{equation}\label{st2.32}
  \textrm{s-GI}_{n}(q,r)=\left(
                    \begin{array}{c}
                     2f_{2n+3}-q g_{2n+2} \\
                      -2h_{2n+3}-r g_{2n+2} \\
                    \end{array}
                  \right)=0,~~n\in\mathbb{N}_0.
\end{equation}
Explicitly,
\begin{align}
  \textrm{s-GI}_0(q,r)=&~\left(
                    \begin{array}{c}
                      -2q_x+2c_1q \\
                      2r_x+2c_1r \\
                    \end{array}
                  \right)=0,\nonumber\\
\textrm{s-GI}_1(q,r)=&~\left(
                    \begin{array}{c}
                      q_{xx}+qq_xr+c_1(-2q_x)+2c_2q \\
                     r_{xx}-qrr_x+2c_1r_x+2c_2r \\
                    \end{array}
                  \right)=0,\nonumber\\
\textrm{s-GI}_2(q,r)=&~\left(
                    \begin{array}{c}
                   -\frac{1}{2}q_{xxx}-\frac{3}{4}qq_{xx}r-\frac{3}{8}q^2r^2q_x-\frac{3}{4}q_x^2r \\
                   \frac{1}{2}r_{xxx}-\frac{3}{4}qrr_{xx}+\frac{3}{8}q^2r^2r_x-\frac{3}{4}qr_x^2\\
                    \end{array}
                  \right)+c_1\left(
                        \begin{array}{c}
                          q_{xx}+qq_xr\\
                          r_{xx}-2qrr_x \\
                        \end{array}
                      \right)\nonumber\\
                  &~~+c_2\left(
                                   \begin{array}{c}
                                     -2q_x \\
                                     2r_x \\
                                   \end{array}
                                 \right)+c_3\left(
                                              \begin{array}{c}
                                                2q \\
                                                2r \\
                                              \end{array}
                                            \right)=0,
                  ~~\textrm{etc.},\nonumber
\end{align}
represent the first few equations of the stationary GI hierarchy.

In the following we shall frequently make the assumption that
$q,r$ satisfy the $n$th stationary GI equations, that is,
they satisfy one of the $n$th stationary GI equations after a particular choice
of integration constants $c_\ell\in\mathbb{C},\ell=1,\ldots,n+1,n\in\mathbb{N}_0.$

Next we introduce polynomials $F_{2n+1},$ $G_{2n+2}$ and $H_{2n+1}$ with respect to
the spectral parameter $z\in\mathbb{C}$,
 \begin{align}
   F_{2n+1}(z)=&~\sum_{\ell=0}^nf_{2\ell+1}z^{2(n-\ell)+1},\label{st2.24}\\
   G_{2n+2}(z)=&\sum_{\ell=0}^{n}g_{2\ell}z^{2(n-\ell)+2}+\frac{1}{2}g_{2n+2},\label{st2.23}\\
   H_{2n+1}(z)=&~\sum_{\ell=0}^nh_{2\ell+1}z^{2(n-\ell)+1},\label{st2.25}
  \end{align}
 and corresponding homogeneous polynomials
 \begin{align}
  &\widehat{F}_0(z)=F_0(z)=2q,~~\nonumber\\
  &\widehat{F}_{2\ell+1}(z)=F_{2\ell+1}(z)|_{c_k=0,k=1,\ldots,\ell}=\sum_{k=0}^\ell\hat{f}_{2k+1}z^{2(\ell-k)+1},~~\ell\in\mathbb{N},\nonumber\\
  &\widehat{G}_0(z)=G_0(z)=2,~~\nonumber\\
  &\widehat{G}_{2\ell+2}(z)=G_{2\ell+2}(z)|_{c_k=0,k=1,\ldots,\ell+1}
  =\sum_{k=0}^{\ell}\hat{g}_{2k}z^{2(n-\ell)+2}+\frac{1}{2}\hat{g}_{2n+2},~~\ell\in\mathbb{N},\nonumber\\
  &\widehat{H}_0(z)=H_0(z)=-2r,~~\nonumber\\
  &\widehat{H}_{2\ell+1}(z)=H_{2\ell+1}(z)|_{c_k=0,k=1,\ldots,\ell}=\sum_{k=0}^\ell\hat{h}_{2k+1}z^{2(\ell-k)+1},~~\ell\in\mathbb{N}.\nonumber
 \end{align}
 Noting (\ref{st2.13})-(\ref{st2.16}), one finds (\ref{st2.28c}), or equivalently, (\ref{st2.32})
 becomes
 \begin{align}
 \widetilde{G}_{2n+2,x}=&~\tilde{r}z\widetilde{F}_{2n+1}+\tilde{q}z\widetilde{H}_{2n+1},\label{st2.36}\\
 \widetilde{F}_{2n+1,x}=&~2\tilde{q}z\widetilde{G}_{2n+2}-(2z^2-\frac{1}{2}\tilde{q}\tilde{r})\widetilde{F}_{2n+1},\label{st2.37}\\
 \widetilde{H}_{2n+1,x}=&~2\tilde{r}z\widetilde{G}_{2n+2}+(2z^2-\frac{1}{2}\tilde{q}\tilde{r})\widetilde{H}_{2n+1},\label{st2.38c}
 \end{align}
 if we identifying $\tilde{q}, \tilde{r}$ with $-q, -r$, respectively.
 Here
 \begin{align}
   \widetilde{F}_{2n+1}(z)=&~\sum_{\ell=0}^n\tilde{f}_{2\ell+1}z^{2(n-\ell)+1}, \\
   \widetilde{G}_{2n+2}(z)=&\sum_{\ell=0}^{n}\tilde{g}_{2\ell}z^{2(n-\ell)+2}+\frac{1}{2}\tilde{g}_{2n+2}, \\
   \widetilde{H}_{2n+1}(z)=&~\sum_{\ell=0}^n\tilde{h}_{2\ell+1}z^{2(n-\ell)+1},
  \end{align}
 and $\{\tilde{f}_{2\ell+1}\}_{\ell=0,\ldots,n}, \{\tilde{g}_{2\ell}\}_{\ell=0,\ldots,n+1},\{\tilde{h}_{2\ell+1}\}_{\ell=0,\ldots,n}$
 satisfy the same recursion
 relations as $\{f_{2\ell+1}\}_{\ell=0,\ldots,n}, \{g_{2\ell}\}_{\ell=0,\ldots,n+1},\{h_{2\ell+1}\}_{\ell=0,\ldots,n}$
 in (\ref{st2.13})-(\ref{st2.16}), that is,
 \begin{align}
  &\tilde{g}_{2\ell+1}=~\tilde{f}_{2\ell}=\tilde{h}_{2\ell}=0,~~\ell\in\mathbb{N}_0, \\
  &\tilde{f}_1=~2\tilde{q},~\tilde{g}_0=2,~\tilde{h}_1=-2\tilde{r},\\
  &2\tilde{f}_{2\ell+1}=-\tilde{f}_{2\ell-1,x}+2\tilde{q}\tilde{g}_{2\ell}+(1/2)\tilde{q}\tilde{r} \tilde{f}_{2\ell-1}, ~\ell\in\mathbb{N}_0,\\
& 2\tilde{h}_{2\ell+1}=~\tilde{h}_{2\ell-1,x}-2\tilde{r}\tilde{g}_{2\ell}+(1/2)\tilde{q}\tilde{r} \tilde{h}_{2\ell-1},~\ell\in\mathbb{N}_0,\\
&\tilde{g}_{2\ell,x}=~\tilde{r}\tilde{f}_{2\ell+1}+\tilde{q}\tilde{h}_{2\ell+1},~\ell\in\mathbb{N}_0.
 \end{align}
 Moreover, (\ref{st2.36})-(\ref{st2.38c}) yield
 \beq
 (\widetilde{G}_{2n+2}^2-\widetilde{F}_{2n+1}\widetilde{H}_{2n+1})_x=0,
 \eeq
  which implies $\widetilde{G}_{2n+2}^2-\widetilde{F}_{2n+1}\widetilde{H}_{2n+1}$ is $x$-independent and hence
  \beq\label{st2.41}
  \widetilde{G}_{2n+2}^2-\widetilde{F}_{2n+1}\widetilde{H}_{2n+1}=R_{4n+4},
  \eeq
  where the integration constant $R_{4n+4}$
  is a polynomial of degree $4n+4$ with respect to $z.$ Let $\{E_m^2\}_{m=0,\ldots,2n+1}$
  be its zeros, then
  \beq
  R_{4n+4}(z)= 4\prod_{m=0}^{2n+1}(z^2-E_m^2),~~\{E_m^2\}_{m=0,\ldots,2n+1}\in\mathbb{C}.
  \eeq
  Next we study the restriction of the differential expression (\ref{st2.28d})
  to the two-dimensional kernel of $M-z^2.$
  More precisely, (\ref{st2.27}) implies
  \beq\label{st2.43}
  Q_{2n+2}|_{\textrm{ker}(M-z^2)}=\left(
                                    \begin{array}{cc}
                                      -G_{2n+2} & F_{2n+1} \\
                                      -H_{2n+2} & G_{2n+2} \\
                                    \end{array}
                                  \right)\Big|_{\textrm{ker}(M-z^2)}.
  \eeq
  We emphasize that the result (\ref{st2.43}) is valid independently
  of whether or not $Q_{2n+2}$ and $M$ commute. However,
  if one makes the additional assumption $Q_{2n+2}$ and $M$ commute, we can prove
  that there exists a algebraic relationship between $Q_{2n+2}$
  and $M$ and this relation will be shown in the following results.

  \newtheorem{thm2.1st}[hyp1]{Theorem}
  \begin{thm2.1st}
  Assume that $Q_{2n+2}$ and $M$ commute, $[Q_{2n+2},M]=0,$ or equivalently, suppose
  $\textrm{s-GI}$$_{n}(q,r)=0$ for some fixed $n\in\mathbb{N}_0.$ Then $M$ and $Q_{2n+2}$
  satisfy an algebraic relationship of type
  \beq\label{st2.44}
  \begin{split}
  \mathcal{F}_n(M,Q_{2n+2})=Q^{2}_{2n+2}-R_{4n+4}(M)=0,\\
  R_{4n+4}=4\prod_{m=0}^{2n+1}(z^2-E_m^2),~~z\in\mathbb{C}.
  \end{split}
  \eeq
  \end{thm2.1st}
 \proof
 From the commutativity of $Q_{2n+2}$ and $M$, the definition of $R_{4n+4}$ (\ref{st2.41})
 and the relations
 \begin{equation*}
 \widetilde{G}_{2n+2}=G_{2n+2},\quad \widetilde{F}_{2n+1}=-F_{2n+1},\quad \widetilde{H}_{2n+1}=-H_{2n+1},
 \end{equation*}
 as well as the expression for $Q_{2n+2}$ on the kernel of $M-z^2$ (\ref{st2.43}), one infers
 \begin{align*}
 Q^{2}_{2n+2}|_{\textrm{ker}(M-z^2)}=&~\left(
                                       \begin{array}{cc}
                                         G_{2n+2}^2-F_{2n+1}H_{2n+1} & 0 \\
                                         0 & G_{2n+2}^2-F_{2n+1}H_{2n+1} \\
                                       \end{array}
                                     \right)\Big|_{\textrm{ker}(M-z^2)}\\
                                     =&~\left(
                                                  \begin{array}{cc}
                                                    R_{4n+4} & 0 \\
                                                    0 & R_{4n+4} \\
                                                  \end{array}
                                                \right)\Big|_{\textrm{ker}(M-z^2)}\\
 =&~R_{4n+4}(M)|_{\textrm{ker}(M-z^2)}.
 \end{align*}
 Thus, $Q_{2n+2}^2$ and $R_{4n+4}$ coincide on $\textrm{ker}(M-z^2).$ Since $z^2\in\mathbb{C}^2$
 is arbitrary, we finally obtain (\ref{st2.44}).
 \qed

 One calls $\mathcal{F}_{n}(M,Q_{2n+2})$ the Burchnall-Chaundy polynomial of the pair
 $[M,Q_{2n+2}].$ The relation (\ref{st2.44}) naturally leads to the hyperelliptic
 curve $\mathcal{K}_n,$
where
\beq\label{st2.45}
\begin{split}
&\mathcal{K}_n:~\mathcal{F}_n(z,y)=y^2-R_{4n+4}(z)=0,\\
&R_{4n+4}(z)=\prod_{m=0}^{2n+1}(z^2-E_m^2),~~\{E_m^2\}_{m=0,\ldots,2n+1}\in\mathbb{C}.
\end{split}
\eeq
Next section we introduce the notations $\eta=z^2, \widetilde{E}_m=E_m^2$, (\ref{st2.45})
changes to the hyperelliptic curve of arithmetic genus $n\in\mathbb{N}_0$ (possibly with a singular affine part),
where
\beq
\begin{split}
&\mathcal{K}_n:~\mathcal{F}_n(\eta,y)=y^2-R_{4n+4}(\eta)=0,\\
&R_{4n+4}(\eta)=\prod_{m=0}^{2n+1}(\eta-\widetilde{E}_m),~~\{\widetilde{E}_m\}_{m=0,\ldots,2n+1}\in\mathbb{C}.
\end{split}
\eeq

Next we introduce the time-dependent GI hierarchy. This means that $q,r$
are now considered an functions of both space and time. For each equation
in the hierarchy, that is, for each $n$, we introduce a deformation parameter $t_n\in\mathbb{R}$
in $q$ and $r,$ replacing $q(x),r(x)$ by $q(x,t_n)$, $r(x,t_n)$. The matrix differential expression
$M$ now becomes
\begin{equation}
  M=\left(
      \begin{array}{cc}
        \partial+\frac{1}{4}q(\cdot,t_n)r(\cdot,t_n) & -zq(\cdot,t_n) \\
        zr(\cdot,t_n) &  -\partial+\frac{1}{4}q(\cdot,t_n)r(\cdot,t_n)\\
      \end{array}
    \right).
\end{equation}
The quantities $\{f_{2\ell+1}\}_{\ell\in\mathbb{N}_0},\{g_{2\ell}\}_{\ell\in\mathbb{N}_0}$
and $\{h_{2\ell+1}\}_{\ell\in\mathbb{N}_0}$
and $Q_{2n+2},n\in\mathbb{N},$ are still defined by
(\ref{st2.13})-(\ref{st2.16}) and (\ref{st2.27}), respectively.
The time-dependent GI hierarchy is obtained by imposing the Lax commutator equations
\beq\label{st2.56m}
\frac{d}{dt_{n}}(-M)-[-Q_{2n+2},-M]=0,~~t_n\in\mathbb{R},
\eeq
or equivalently,
\begin{align}\label{st2.39c}
\left(
  \begin{array}{cc}
    -\frac{1}{4}(qr)_{t_n}+\frac{1}{2}g_{2n+2,x} &  q_{t_n}-2f_{2n+3}+q g_{2n+2}\\
    -r_{t_n}+2h_{2n+3}+r g_{2n+2} &  \frac{1}{4}(qr)_{t_n}+\frac{1}{2}g_{2n+2,x} \\
  \end{array}
\right)=0.
\end{align}
From (\ref{st2.39c}), we obtain
\begin{align*}
\frac{1}{4}(qr)_{t_n}=&~\frac{1}{4}(rq_{t_n}+qr_{t_n})\\
=&~\frac{1}{4}[r(2f_{2n+3}-qg_{2n+2})+q(2h_{2n+3}+rg_{2n+2})]\\
=&~\frac{1}{2}g_{2n+2,x},
\end{align*}
that is,
\beq
\frac{1}{2}g_{2n+2,x}-\frac{1}{4}(qr)_{t_n}=0.
\eeq
Thus, varying $n$, the time-dependent GI hierarchy can be written as
\beq\label{st2.59a}
\textrm{GI}_{n}(q,r)=\left(
                     \begin{array}{c}
                       q_{t_n}-2f_{2n+3}+qg_{2n+2} \\
                       r_{t_n}-2h_{2n+3}-rg_{2n+2} \\
                     \end{array}
                   \right)=0,\quad t_n\in\mathbb{R},\quad n\in\mathbb{N}_0.
\eeq
Explicitly,
\begin{align}
  \textrm{GI}_0(q,r)=&~\left(
                    \begin{array}{c}
                      q_{t_0}+2q_x-2c_1q \\
                      r_{t_0}+2r_x+2c_1r \\
                    \end{array}
                  \right)=0,\nonumber\\
\textrm{GI}_1(q,r)=&~\left(
                    \begin{array}{c}
                    q_{t_1}-q_{xx}-qq_xr+2c_1q_x-2c_2q \\
                    r_{t_1}+r_{xx}-qrr_x+2c_1r_x+2c_2r \\
                    \end{array}
                  \right)=0,\nonumber\\
\textrm{GI}_2(q,r)=&~\left(
                    \begin{array}{c}
                   q_{t_2}+\frac{1}{2}q_{xxx}+\frac{3}{4}qq_{xx}r+\frac{3}{8}q^2r^2q_x+\frac{3}{4}q_x^2r \\
                   r_{t_2}+\frac{1}{2}r_{xxx}-\frac{3}{4}qrr_{xx}+\frac{3}{8}q^2r^2r_x-\frac{3}{4}qr_x^2\\
                    \end{array}
                  \right)+c_1\left(
                        \begin{array}{c}
                          -q_{xx}-qq_xr\\
                          r_{xx}-2qrr_x \\
                        \end{array}
                      \right)\nonumber\\
                  &~~+c_2\left(
                                   \begin{array}{c}
                                     2q_x \\
                                     2r_x \\
                                   \end{array}
                                 \right)+c_3\left(
                                              \begin{array}{c}
                                                -2q \\
                                                2r \\
                                              \end{array}
                                            \right)=0,
                  ~~\textrm{etc.},\nonumber
\end{align}
are the first few equations of the time-dependent GI hierarchy.
The system of equation $\textrm{GI}_1(q,r)=0$ with $c_1=c_2=0$
represents the GI system.

\newtheorem{zerocur}[hyp1]{Remark}
\begin{zerocur}
An alternative construction of the GI hierarchy can be introduced by developing its zero-curvature formulism.
One defines
 \begin{align}
   U(z)=&~
    \left(
      \begin{array}{cc}
        -z^2+\frac{1}{4}qr & qz \\
        rz & z^2-\frac{1}{4}qr \\
      \end{array}
    \right),~~z\in\mathbb{C},\label{2.1}~~\\
    V_{2n+2}(z)=&
    \left(
      \begin{array}{cc}
        -G_{2n+2}(z) & F_{2n+1}(z) \\
        -H_{2n+2}(z) & G_{2n+2}(z) \\
      \end{array}
    \right),\quad n\in\mathbb{N}_0,\label{2.2}
 \end{align}
where
 $G_{n+1},F_n,H_n$ are polynomials defined in (\ref{st2.24})-(\ref{st2.25}).
 Hence, the stationary part of this section, being a consequence of $[Q_{2n+2},M]=0$, can equivalently
be based on the stationary zero-curvature equation
\begin{align}
0=&~-V_{2n+2,x}(z)+[U(z), V_{2n+2}(z)]\nonumber\\
=&~\left(
     \begin{smallmatrix}
        G_{2n+2,x}-zr F_{2n+1}-q zH_{2n+1} &  -F_{2n+1,x}+2zq G_{2n+2}+(-2z^2+\frac{1}{2} q r) F_{2n+1} \\
        H_{2n+1,x}-2rz G_{2n+2}+(-2z^2+\frac{1}{2}qr)H_{2n+1} &-G_{2n+2,x}+z r F_{2n+1}+q zH_{2n+1}  \\
     \end{smallmatrix}
   \right)\nonumber\\
   =&~\left(
        \begin{array}{cc}
          \frac{1}{2}g_{2n+2,x} & (-qg_{2n+2}+2f_{2n+3})z \\
          (rg_{2n+2}+2 h_{2n+3})z & -\frac{1}{2}g_{2n+2,x}\\
        \end{array}
      \right).
\end{align}
In particular, the hyperelliptic curve $\mathcal{K}_n$ can also be obtained from
the characteristic equation of $V_{2n+2}$ by
\begin{align}
\textrm{det}(yI_2-V_{2n+2}(z))=&~y^2-\textrm{det}(V_{2n+2})\nonumber\\
=&~y^2-G_{2n+2}^2(z)+F_{2n+1}(z)H_{2n+1}(z)\nonumber\\
=&~y^2-\mathscr{R}_{4n+4}(z)=0.
\end{align}
Similarly, the time-dependent part (\ref{st2.56m}), (\ref{st2.39c}), (\ref{st2.59a}),
being based on the Lax equation (\ref{st2.56m}), can equivalently developed from
the zero-curvature equation
\begin{align}
0=&~U_{t_n}-V_{2n+2,x}+[U,V_{2n+2}]\nonumber\\
=&~\left(
     \begin{smallmatrix}
        (qr)_{t_n}+G_{2n+2,x}-zr F_{2n+1}-q zH_{2n+1} &  q_{t_n}z-F_{2n+1,x}+2zq G_{2n+2}+(-2z^2+\frac{1}{2} q r) F_{2n+1} \\
        r_{t_n}z+H_{2n+1,x}-2rz G_{2n+2}+(-2z^2+\frac{1}{2}qr)H_{2n+1} &-(qr)_{t_n}-G_{2n+2,x}+z r F_{2n+1}+q zH_{2n+1}  \\
     \end{smallmatrix}
   \right)\nonumber\\
   =&~\left(
        \begin{array}{cc}
          (qr)_{t_n}+\frac{1}{2}g_{2n+2,x} & (q_{t_n}-qg_{2n+2}+2f_{2n+3})z \\
          (r_{t_n}+rg_{2n+2}+2 h_{2n+3})z & -(qr)_{t_n}-\frac{1}{2}g_{2n+2,x}\\
        \end{array}
      \right).
\end{align}

\end{zerocur}\vspace{0.3cm}

Relations between the mKP hierarchy and the GI hierarchy can be
summarized as follows.

\newtheorem{thmnew}{Theorem}[section]
\begin{thmnew}[see \cite{cheng1992m} or \cite{Konopelchenko1992a}]
Assume $q,r$ is a compatible solution of the system
 \begin{align}
    &\widehat{\textrm{GI}}_{1}(q,r)=0, \label{st2.38}\\
    &\widehat{\textrm{GI}}_{p}(q,r)=0, ~~p\geq 2,\label{st2.39}
 \end{align}
 where we denote by
 \begin{equation}
  \widehat{\textrm{GI}}_{n}(q,r)=\textrm{GI}_{n}(q,r)|_{c_k=0,k=1,\ldots,n+1},~~n\in\mathbb{N}_0,
\end{equation}
the corresponding homogeneous GI equations.
 Then
 $$u(x,y,t_{p})=-2^{p-1}q(-2x, y,(-2)^{p-1}t_{p})r(-2x, y,(-2)^{p-1}t_{p})$$
yields a solution of the $p$th KP equation.
\end{thmnew}

Given these preparations, next we shall study the algebro-geometric solutions of the KP hierarachy.
Now $q,r$ are considered as functions of variables $x,t_1,t_{p}$ with $t_1=y$. Then
we introduce the following auxiliary linear problem \footnote{One can also start from the
following linear problems
 \begin{equation}\label{2.143433}
\begin{split}
  &\psi_{x}(z)=U(z)\psi(z), \\
  &\psi_{t_{m}}(z)=\widehat{V}_{2m+2}(z)\psi(z),~~z\in\mathbb{C},~~m=1,2,3,\ldots
  \end{split}
\end{equation}
where $q,r$ are considered as functions of $x,y,t_2,\ldots$
and obtain similar results.
}
\begin{equation}\label{2.14k}
\begin{split}
  &\psi_{x}(z)=U(z)\psi(z), \\
  &\psi_{y}(z)=\widehat{V}_4(z)\psi(z),\\
  &\psi_{t_{2}}(z)=\widehat{V}_{2p+2}(z)\psi(z),~~z\in\mathbb{C},~~p\geq 2,
  \end{split}
\end{equation}
where $$\widehat{V}_{2k+2}(z)=V_{2k+2}(z)|_{c_{\ell}=0,\ell=1,\ldots,k+1},$$
and
$\psi(z)=(\psi_1(z,x,y,t_{p}),\psi_2(z,x,y,t_{p}))^T.$
Let
$$\psi^\pm(z)=(\psi_1^{\pm}(z,x,y,t_{p}),\psi_2^{\pm}(z,x,y,t_{p}))$$
 be two fundamental solutions of linear system (\ref{2.14k}) and three squared basis functions $\mathscr{G}, \mathscr{F}, \mathscr{H}$ can be defined in term of $\psi^\pm_1(z), \psi^{\pm}_2(z)$ by
\begin{equation}\label{2.6}
\begin{split}
  \mathscr{G}(z,x,y,t_p)=&~ \frac{1}{2}(\psi_1^+(z,x,y,t_p)\psi_2^-(z,x,y,t_p)\nonumber\\
  &~+\psi_1^-(z,x,y,t_p)\psi_2^+(z,x,y,t_p)), \\
  \mathscr{F}(z,x,y,t_p)=&~\psi_1^+(z,x,y,t_p)\psi_1^-(z,x,y,t_p),\\
  \mathscr{H}(z,x,y,t_p)=&~\psi_2^+(z,x,y,t_p)\psi_2^-(z,x,y,t_p).
\end{split}
\end{equation}
Using
(\ref{2.1}), (\ref{2.2}) and (\ref{2.14k}), we derive
the following linear system
\begin{align}
 & \mathscr{G}_{x}=rz \mathscr{F}+qz\mathscr{H},\label{2.16k} \\
 & \mathscr{G}_{y}=\widehat{F}_{3}\mathscr{H}-\widehat{H}_3 \mathscr{F},\\
 & \mathscr{G}_{t_{p}}=\widehat{F}_{2p+1}\mathscr{H}-\widehat{H}_{2p+1} \mathscr{F},\\
 & \mathscr{F}_{x}=2qz\mathscr{G}+\left(-2z^2+\frac{1}{2}qr\right)\mathscr{F},\label{2.19k}\\
 &\mathscr{F}_{y}=2\widehat{F}_3\mathscr{G}-2\widehat{G}_4\mathscr{F},\label{2.20kp}\\
 &\mathscr{F}_{t_{p}}=2\widehat{F}_{2p+1}\mathscr{G}-2\widehat{G}_{2p+2}\mathscr{F},\label{2.21kp}\\
 &\mathscr{H}_{x}=2rz\mathscr{G}+\left(2z^2-\frac{1}{2}qr\right)\mathscr{H},\label{2.22k}\\
 &\mathscr{H}_{y}=2\widehat{G}_4\mathscr{H}-2\widehat{H}_3\mathscr{G},\label{2.23k}\\
 & \mathscr{H}_{t_{p}}=2\widehat{G}_{2p+2}\mathscr{H}-2\widehat{H}_{2p+1}\mathscr{G}.\label{2.24k}
\end{align}
For each fixed $p$, solutions of linear system (\ref{2.16k})-(\ref{2.24k}) are connected with the following
basic initial value problem of GI system (\ref{st2.51}).

\newtheorem{th1}{Theorem}[section]
\begin{th1}
Assume $q,r\in C^\infty(\mathbb{R}^{2+1})$.
Moreover, if $(q,r)$ solves (\ref{st2.38}), (\ref{st2.39}),
then the set
\begin{equation}\label{2.29k}
\mathcal{E}=\textrm{span}\{(\mathscr{G},\mathscr{F},\mathscr{H})|~
(G_{2j+2},F_{2j+1},H_{2j+1}),~ j\in\mathbb{N}_0\}
\end{equation}
forms a vector space on $\mathbb{C}$
for solutions of the system
(\ref{2.16k})-(\ref{2.24k}). In particular, the functions $q,r$ which constituting
$(G_{2n+2},$ $F_{2n+1},H_{2n+1})$, $n\in\mathbb{N}_0,$
are connected with solutions of the following initial problem
 \begin{align}\label{st2.51}
\begin{cases}
q_{y}=-2f_{5}+qg_{4},\\
r_{y}=-2 h_{5}-rg_{4},\\
q_{t_{p}}=-2f_{2p+3}+qg_{2p+2},\\
r_{t_{p}}=-rg_{2p+2}-2 h_{2p+3},\\
 2f_{2n+3}-qg_{2n+2}=0,\\
 rg_{2n+2}+2 h_{2n+3}=0,~~ p\geq 2.
 \end{cases}
\end{align}

\end{th1}
\proof
It is not difficult to verify that
(\ref{st2.51}) is equivalent to the following zero
curvature representation
\begin{equation}\label{2.9k}
\begin{split}
  &\frac{\partial}{\partial x}V_{2j+2}(z)-\frac{\partial}{\partial t_j}U(z)=[U(z),V_{2j+2}(z)],~j=2,p,n,
\end{split}
\end{equation}
from remark 2.3.
Next we shall show
\begin{equation}\label{2.30k}
\begin{split}
  &[V_{2j+2}(z)+\frac{\partial}{\partial t_j},V_{2k+2}(z)+\frac{\partial}{\partial t_k}]
  =0,~j,k=2,p,\\
  &[V_{2j+2}(z)+\frac{\partial}{\partial t_j}, V_{2n+2}(z)]=0,~j=2,
  \end{split}
\end{equation}
which can be proved as follows. First,
taking into account the $2\times 2$
matrix-valued differential expression
 \begin{align}
  M=&~\left(
      \begin{array}{cc}
        \partial+\frac{1}{4}qr & -zq \\
        zr &  -\partial+\frac{1}{4}qr\\
      \end{array}
    \right),
                \\
 Q_{2n+2}=&~\sum_{j=0}^n\left(
             \begin{array}{cc}
               -g_{2\ell}z^2 & f_{2\ell+1}z \\
               -h_{2\ell+1}z & g_{2\ell+2}z^2 \\
             \end{array}
           \right)M^{n-\ell}
           +\left(
              \begin{array}{cc}
                \frac{1}{2}g_{2n+2} & 0 \\
                0 & -\frac{1}{2}g_{2n+2} \\
              \end{array}
            \right),\nonumber\\
            &~~~~~~~~~~~~~~~~~~~~~~~~~~~~~~~~~~~~~~~~~n\in\mathbb{N}_0,~f_{-1}=h_{-1}=0,
 \end{align}
one infers (\ref{2.9k}) is also equivalent to
\begin{equation*}
\begin{split}
  &[Q_{2j+2}+\frac{\partial}{\partial t_j},M]=0,~~j=2,~p,\\
  & [Q_{2n+2},M]=0.
\end{split}
\end{equation*}
Then by Corollary 2 of Theorem 4.2 in \cite{Krichever} it follows
 \begin{align*}
  &[Q_{2j+2}+\frac{\partial}{\partial t_j},Q_{2k+2}+\frac{\partial}{\partial t_k}]=0,~~j,~k=2, p, \\
  &[Q_{2j+2}+\frac{\partial}{\partial t_j},Q_{2n+2}]=0,~~j=2, p,
\end{align*}
and hence
\begin{align}
  [V_{2j+2}(z)+\frac{\partial}{\partial t_j},V_{2k+2}(z)+\frac{\partial}{\partial t_k}]&=[Q_{2j+2}+\frac{\partial}{\partial t_j},Q_{k+1}+\frac{\partial}{\partial t_k}]|_{\textrm{ker}(M-z)}\nonumber\\
  &=0,~~j,~k=2,p, \label{2.86xy} \\
   [V_{2j+2}(z)+\frac{\partial}{\partial t_j},V_{2n+2}(z)]&=[Q_{2j+2}+\frac{\partial}{\partial t_j},Q_{2n+2}]|_{\textrm{ker}(M-z)}\nonumber\\
  &=0,~~j=2,p,
\end{align}
holds,
where
\begin{equation*}
  \textrm{ker}(M-z)=\{\Psi=\left(
                             \begin{array}{c}
                               \psi_1 \\
                               \psi_2 \\
                             \end{array}
                           \right):\mathbb{R}^{2+1}\rightarrow \mathbb{C}_{\infty}^2|(M-z)\Psi=0
  \},~z\in\mathbb{C}.\textrm{\qed}
\end{equation*}

\newtheorem{rem1}[th1]{Remark}
\begin{rem1}
This proposition can also be proved by introducing a meromorphic function $\phi$ and
considering $q,r$ as functions of three variables $x,y,t_{p}$ (see \cite{yfz}, Lemma 5.3).

\end{rem1}
\section{Spectral Curve, Dubrovin-type Equations and Trace Formula}
In the following, we shall introduce the spectral curve associated with
mKP hierarchy. To emphasize the difference between different solutions in (\ref{2.29k}),
we add a subscript in each of $\mathscr{G},\mathscr{F},\mathscr{H}$, that is,
\begin{equation*}
(\mathscr{G}_{2n+2},\mathscr{F}_{2n+1},\mathscr{H}_{2n+1})=(G_{2n+2},F_{2n+1},H_{2n+1}),~n\in\mathbb{N}. \end{equation*}
 Then using (\ref{2.16k})-(\ref{2.24k}) and theorem 2.1, we have
\begin{equation}
(\mathscr{G}^2_{2n+2}-\mathscr{F}_{2n+1}\mathscr{H}_{2n+1})_{t_j}=0,~~n\in\mathbb{N},~j=1,2,p,
\end{equation}
and hence $\mathscr{G}^2_{n+1}-\mathscr{F}_{n}\mathscr{H}_{n}$ is $x,y,t_{p}$-independent implying
\begin{equation}\label{3.3zp}
\mathscr{G}^2_{2n+2}-\mathscr{F}_{2n+1}\mathscr{H}_{2n+1}=\mathscr{R}_{4n+4},~~n\in\mathbb{N},
\end{equation}
where integration constant $\mathscr{R}_{4n+4}$ is a polynomial of degree $4n+4$
in $z$.
 Let
\begin{equation}\label{2.21}
\begin{split}
  \mathscr{R}_{4n+4}&(z)=\sum_{j=0}^{2n+2-j}(-1)^js_j z^{2n+2-j},\\
  &s_0=4,~s_{j}\in\mathbb{C},j=1,\ldots,2n,
\end{split}
\end{equation}
and hyperelliptic curve associated with the $p$th mKP equation is then introduced
as follows:
\begin{equation}\label{2.22}
\begin{split}
  X: \mathscr{P}(\eta,y)=&~y^2-\frac{1}{4}\mathscr{R}_{4n+4}(\sqrt{\eta}) \\
  =&~y^2-\frac{1}{4}\mathscr{G}_{2n+2}^2(\sqrt{\eta})+\frac{1}{4}\mathscr{F}_{2n+1}(\sqrt{\eta})\mathscr{H}_{2n+1}(\sqrt{\eta})\\
  =&~0.
  \end{split}
\end{equation}
The curve $X$ is compactified by joining two points $P_{\infty\pm}$ at infinity but for notational simplicity the compactification is also denoted by $X$. Points $P$ on $X\backslash \{P_{\infty\pm}\}$ are denote by pairs $(\eta,y)$, where $y(\cdot)$ is the
meromorphic function on $X$ satisfying $\mathscr{P}(\eta,y)=0.$ The complex structure
on $X$ is defined in the usual way. Hence, $X$ becomes
a two-sheeted hyperelliptic Riemann surface of (arithmetic) genus $n$ in a standard manner.
Moreover, we denote the upper and lower sheets $\Pi_\pm$ by
\begin{equation*}
  \Pi_\pm=\{(\eta,\pm\frac{1}{2}\sqrt{\mathscr{R}_{4n+4}(\sqrt{\eta})})\in X|\eta\in \Pi\},
\end{equation*}
where $\Pi$ denotes the cut plane $\mathbb{C}\backslash\mathcal{C}$
and $\mathcal{C}$ is the union of $n$ nonintersecting cuts joining two different
branches of $\sqrt{\mathscr{R}_{4n+4}(\sqrt{\eta})}$. The holomorphic sheet exchange map on $X$ is defined by
\begin{align*}
&*:\quad X\rightarrow X, \\
&P=(\eta,\frac{1}{2}\sqrt{\mathscr{R}_{4n+4}(\eta)})\mapsto P^{*}=(\eta,-\frac{1}{2}\sqrt{\mathscr{R}_{4n+4}(\sqrt{\eta})}),\quad\\
&P_{\infty\pm}\mapsto P_{\infty\pm}^{*}=P_{\infty\mp}.
\end{align*}
 Moreover, positive divisors on $X$ of degree $n$
   are denoted by
        \begin{equation}\label{3.4xyz}
          \mathcal{D}_{P_1,\ldots,P_{n}}:
             \begin{cases}
              X\rightarrow \mathbb{N}_0,\\
              P\rightarrow \mathcal{D}_{P_1,\ldots,P_{n}}(P)=
                \begin{cases}
                  \textrm{ $k$ if $P$ occurs $k$
                      times in $\{P_1,\ldots,P_{n}\},$}\\
                   \textrm{ $0$ if $P \notin
                     $$ \{P_1,\ldots,P_{n}\}.$}
                \end{cases}
             \end{cases}
        \end{equation}
        In particular, the divisor $\left(\phi(\cdot)\right)$ of a meromorphic function
        $\phi(\cdot)$ on X is defined by
        \begin{eqnarray}\label{3.4a0xyz}
          \left(\phi(\cdot)\right):
          X \rightarrow \mathbb{Z},\quad
          P\mapsto \omega_{\phi}(P),
          \end{eqnarray}
        where $\omega_{\phi}(P)=m_0\in\mathbb{Z}$ if
        $(\phi\circ\zeta_{P}^{-1})(\zeta)=\sum_{n=m_0}^{\infty}c_n(P)\zeta^n$
        for some $m_0\in\mathbb{Z}$ by using a chart $(U_{P}, \zeta_P)$ near $P\in X.$
        Finally, we introduce symmetric functions of $x_1,\ldots,x_n$
\begin{equation}\label{2.25}
\begin{split}
& \Psi_{0}(\underline{x})=1,~~\Psi_{1}(\underline{x})=\sum_{j=1}^{n}x_j,~~\Psi_{2}(\underline{x})
=\sum_{ j,k=1,j<k
}^{n}x_jx_k, \\
&\Psi_{3}(\underline{x})
=\sum\nolimits_{  j,k,\ell=1,j<k<\ell
 }^{n}x_jx_kx_\ell, ~~\textrm{etc.,}
 \end{split}
\end{equation}
 where
 $$\underline{x}=(x_1,\ldots,x_{n}).$$

Now we turn to the parameter representations of $\mathscr{G}_{2n+2}, \mathscr{F}_{2n+1}, \mathscr{H}_{2n+1}$,
which are described by the evolution of auxiliary spectrum points $\mu_j(x,y,t_{p}),$ \linebreak $\nu_j(x,y,t_{p}),j=1,\ldots,n$. This procedure is standard, which is similar with 1+1 dimensional case.

\newtheorem{th2}{Theorem}[section]
\begin{th2}
Solutions of (\ref{2.16k})-(\ref{2.24k}) can also be expressed as
\begin{equation}\label{2.23}
  \begin{split}
  &\mathscr{G}_{2n+2}=\sum_{j=0}^{n+1} (-1)^j \dot{g}_{2j}(x,y,t_{p})z^{2n+2-2j},~~ \\
 & \mathscr{F}_{2n+1}=2q(x,y,t_{p})z\prod_{j=1}^{n} (z^2-\mu_j(x,y,t_{p})),\\
 & \mathscr{H}_{2n+1}=-2r(x,y,t_{p})z\prod_{j=1}^{n} (z^2-\nu_j(x,y,t_{p})),
  \end{split}
\end{equation}
where
\begin{align}\label{2.24}
  \dot{g}_0=&~2,~\nonumber\\
  \dot{g}_2=&~qr+\frac{s_1}{4},\\
  \dot{g}_{2j}=&~\frac{1}{4}\Big[-\sum_{\nu=1}^{j-1}\dot{g}_\nu \dot{g}_{j-\nu}-4qr \sum_{\alpha+\beta=j}\Psi_{\alpha}(\underline{\mu})\Psi_{j-\beta}(\underline{\nu})+s_j\Big],~s=2,\ldots,n+1,\nonumber
\end{align}
and $\{\mu_j(x,y,t_{p})\}_{j=1}^n, \{\nu_j(x,y,t_{p})\}_{j=1}^n$ are $n$ roots of $\mathscr{F}_{2n+1},\mathscr{H}_{2n+1}$, respectively. Moreover, if $\{\mu_j(x,y,t_{p})\}_{j=1}^n$ are mutually distinct and finite, then
 they satisfy the
 Dubrovin-type equations
 \begin{align}
    &\mu_{j,x}=-\frac{\sqrt{\mu_j\mathscr{R}_{4n+4}(\sqrt{\mu_j})}}{ \prod_{k\neq j}(\mu_j-\mu_k)},\label{2.28l}\\
    &\mu_{j,y}=-\frac{\widehat{F}_5(\sqrt{\mu_j})\sqrt{\mathscr{R}_{4n+4}(\sqrt{\mu_j})}}{q \prod_{k\neq j}(\mu_j-\mu_k)},\label{2.29m}\\
    &\mu_{j,t_{p}}=-\frac{\widehat{F}_{2p+1}(\sqrt{\mu_j})\sqrt{\mathscr{R}_{4n+4}(\sqrt{\mu_j})}}{q \prod_{k\neq j}(\mu_j-\mu_k)},\label{2.29n}
 \end{align}
 and similar statement is also true for  $\{\nu_j(x,y,t_{p})\}_{j=1}^n$, where (\ref{2.28l})-(\ref{2.29n})
 change to
 \begin{align}
    &\nu_{j,x}=-\frac{\sqrt{\mathscr{R}_{4n+4}(\sqrt{\nu_j})}}{ \prod_{k\neq j}(\nu_j-\nu_k)},\label{2.30}\\
    &\nu_{j,y}=-\frac{\widehat{H}_5(\sqrt{\mu_j})\sqrt{\mathscr{R}_{4n+4}(\sqrt{\mu_j})}}{r \prod_{k\neq j}(\mu_j-\mu_k)},\label{2.30m}\\
    &\nu_{j,t_{p}}=-\frac{\widehat{H}_{2p+1}(\sqrt{\mu_j})\sqrt{\mathscr{R}_{4n+4}(\sqrt{\mu_j})}}{r \prod_{k\neq j}(\mu_j-\mu_k)}.\label{2.30n}
 \end{align}
Finally, $q,r$ and $\mu_j,\nu_j$ are connected by the following trace formula
 \begin{equation}\label{2.32}
 \begin{split}
   &\sum_{j=1}^{n}\mu_j=\frac{1}{4}qr+\frac{1}{2}\frac{q_x}{q}+\frac{s_1}{8}, \\
   &\sum_{j=1}^{n}\nu_j=\frac{1}{4}qr-\frac{1}{2}\frac{r_x}{r}+\frac{s_1}{8}.
 \end{split}
 \end{equation}
\end{th2}
\proof First, insertion of (\ref{2.23}) into (\ref{3.3zp}), (\ref{2.21}) and a comparison powers of $z$ yields (\ref{2.24}). Then by (\ref{2.19k}), (\ref{2.20kp}), (\ref{2.21kp}), (\ref{2.22k}), (\ref{2.23k})
and (\ref{2.24k}),
taking into account (\ref{2.22}), (\ref{2.23}),
one derives (\ref{2.28l})-(\ref{2.30n}). Moreover, combining (\ref{st2.23}) with (\ref{2.24}) yields
\begin{equation}\label{2.26}
\begin{split}
   c_1=-\frac{1}{8} s_1.
\end{split}
\end{equation}
Finally, the formula (\ref{2.32}) is the direct result of (\ref{2.23}), (\ref{2.26}) and theorem 2.1.
\qed

\section{Baker-Akhiezer Function}
It is well known that the Baker-Akhiezer function plays a very important role in finite gap integration of
soliton equations and it permits us to obtain the Riemann theta function representation for solutions of a
given equation and there are numerous articles have been
devoted to this subject \cite{Belokolos1994},~\cite{gesztesy1999},~\cite{gesztesy2003a},~\cite{Kamchatnov}.
This section aims to provide the explicit form of "Baker-Akhiezer function" associated with the $p$th mKP equation, which consists of spectral parameter $z$, potentials $q,r$,
and then study its analytic properties.
Moreover, we will find the conservation relations of soliton equations play a key role in the construction of Baker-Akhiezer function, which reveals symmetry is the intrinsic character of classical integrable system.

Let us first study the Baker-Akhiezer function
$$\psi(P)=(\psi_1(P,x,y,t_{p}),\psi_2(P,x,y,t_{p})),~P\in X,$$
which satisfies
\begin{equation}\label{4.1k}
\begin{split}
  &\psi_x(P)=~U(z)\psi(P),~
  \psi_y(P)=~\widehat{V}_4(z)\psi(P),\\
  &\psi_{t_{p}}(P)=~\widehat{V}_{2p+2}(z)\psi(P),~~p>2,\\
  &\psi_1(P,x_0,y_0,t_{p,0})=1,~~(P,x_0,y_0,t_{p,0})\in X\times\mathbb{R}^3.
  \end{split}
\end{equation}
To find its explicit form, we need some preparations.

\newtheorem{lem4.1}{Lemma}[section]
\begin{lem4.1}
Suppose $q,r\in C^{\infty}(\mathbb{R}^{2+1})$ and $z\in\mathbb{C}$.
Then
$\mathscr{G}_{2n+2}, \mathscr{F}_{2n+1},
\mathscr{H}_{2n+1}$ and basic fundamental solutions $(\psi_1^\pm,\psi_2^\pm)$ of linear system (\ref{4.1k})
have the following algebraic relation:
\begin{align}
 &(\psi_1^+(z)\psi_2^-(z)-\psi_1^-(z)\psi_2^+(z))^2=4 \mathscr{R}_{4n+4}^2(z).\label{4.2k}
 \end{align}
 If we take $\psi_1^+(z)\psi_2^-(z)-\psi_1^-(z)\psi_2^+(z)=2\sqrt{\mathscr{R}_{4n+4}(z)},$
 then
 \begin{align}
 &\psi_1^+(z)\psi_2^-(z)=\mathscr{G}_{2n+2}(z)+\sqrt{\mathscr{R}_{4n+4}(z)},\\
 &\psi_1^-(z)\psi_2^+(z)=\mathscr{G}_{2n+2}(z)-\sqrt{\mathscr{R}_{4n+4}(z)},\\
 &\frac{\psi_2^\pm(z)}{\psi_1^\pm(z)}=\frac{\mathscr{H}_{2n+1}(z)}{\mathscr{G}_{2n+2}(z)\pm \sqrt{\mathscr{R}_{4n+4}(z)}}=\frac{\mathscr{G}_{2n+2}\mp \sqrt{\mathscr{R}_{4n+4}(z)}}{\mathscr{F}_{2n+1}}.\label{4.5k}
\end{align}
\end{lem4.1}
\proof Expressions (\ref{4.2k})-(\ref{4.5k}) can easily be verified by (\ref{2.6}), (\ref{3.3zp}). \qed

\newtheorem{lem4.2}[lem4.1]{Lemma}
\begin{lem4.2}
Suppose $q,r\in C^{\infty}(\mathbb{R}^{2+1})$. Then we have the following relations
\begin{align}
 &\Big[\frac{q(x,y,t_{p})z}{\mathscr{F}_{2n+1}(z,x,y,t_{p})}\Big]_{t_{j}}
 =\Big[\frac{{\widehat{F}_{2j+1}(z,x,y,t_{p})}}{\mathscr{F}_{2n+1}(z,x,y,t_{p})}\Big]_{x},\\
 &
 \Big[\frac{r(x,y,t_{p})z}{\mathscr{H}_{2n+1}(z,x,y,t_{p})}\Big]_{t_{j}}=
 -\Big[\frac{\widehat{H}_{2j+1}(z,x,y,t_{p})}{\mathscr{F}_{2n+1}(z,x,y,t_{p})}\Big]_{x},~~j=2,p,\label{4.6kp}\\
 &\Big[\frac{{\widehat{F}_3(z,x,y,t_{p})}}{\mathscr{F}_{2n+1}(z,x,y,t_{p})}\Big]_{t_{p}}
 =\Big[\frac{{\widehat{F}_{2p+1}(z,x,y,t_{p})}}{\mathscr{F}_{2n+1}(z,x,y,t_{p})}\Big]_{y}, \\
  &\Big[\frac{{\widehat{H}_3(z,x,y,t_{p})}}{\mathscr{F}_{2n+1}(z,x,y,t_{p})}\Big]_{t_{p}}
  =\Big[\frac{{\widehat{H}_{2p+1}(z,x,y,t_{p})}}{\mathscr{F}_{2n+1}(z,x,y,t_{p})}\Big]_{y}.\label{4.8kp}
\end{align}
\end{lem4.2}
\proof The proof is straightforward. By (\ref{2.19k}), (\ref{2.20kp}), (\ref{2.21kp}) and
\begin{equation}
  q_{t_j}-F_{2j+1,x}+2zq G_{2j+2}+(-2z^2+\frac{1}{2} q r) F_{2j+1}=0,~~j=2,p,
\end{equation}
we have
\begin{equation}
\begin{split}
  \Big[\frac{qz}{\mathscr{F}_{2n+1}}\Big]_{t_j}=&~\frac{q_{t_j}z}{\mathscr{F}_{2n+1}}-
  \frac{qz \mathscr{F}_{2n+1,t_j}}{\mathscr{F}_{2n+1}^2}\nonumber\\
  =&~\frac{\widehat{F}_{2j+1,x}+(2z^2-\frac{1}{2}qr)
  \widehat{F}_{2j+1}-2qz\widehat{G}_{2j+2}}{\mathscr{F}_{2n+1}}-\frac{qz(2\widehat{F}_{2j+1}
  \mathscr{G}_{2n+2}-2\widehat{G}_{2j+2}
  \mathscr{F}_{2n+1})}{\mathscr{F}_{2n+1}^2}\nonumber\\
  =&\frac{\widehat{F}_{2j+1,x}+\frac{(\mathscr{F}_{2n+1,x}-2qzG_{2n+2})}{\mathscr{F}_{2n+1}}\widehat{F}_{2j+1}}{\mathscr{F}_{2n+1}}-\frac{2qz\widehat{F}_{2j+1}
  \mathscr{G}_{2n+2}}{\mathscr{F}_{2n+1}^2}\nonumber\\
  =&\frac{\widehat{F}_{2j+1,x}\mathscr{F}_{2n+1}-\widehat{F}_{2j+1}\mathscr{F}_{2n+1,x}}{\mathscr{F}_{2n+1}^2}= \Big[\frac{{\widehat{F}_{2j+1}}}{\mathscr{F}_{2n+1}}\Big]_{x},~~j=2,p.\nonumber\\
  \end{split}
\end{equation}
Moreover, (\ref{4.6kp})-(\ref{4.8kp}) follows by the relations
\begin{equation}
  r_{t_n}z+\widehat{H}_{2n+1,x}-2zr\widehat{G}_{2n+2}+(-2z^2+\frac{1}{2}qr)\widehat{H}_{2n+1}=0,
\end{equation}
(\ref{2.86xy}), (\ref{2.22k}), (\ref{2.23k}) and (\ref{2.24k}).
\qed\vspace{0.3cm}

Now we lift the fundamental solution $(\psi_1^\pm(z),\psi_2^\pm(z))$, $z\in\mathbb{C},$ to Riemann surface $X$.
Define the Baker-Akhiezer function by
\begin{align*}
&\psi_j(P,x,y,t_{p})=  \psi_j^+(z,x,y,t_{p}),~~ j=1,2,~~ y(P)=\frac{1}{2}\sqrt{\mathscr{R}_{4n+4}(z)}, ~~\textrm{for}~~ P\in \Pi_+, \\
& \psi_j(P,x,y,t_{p})= \psi_j^-(z,x,y,t_{p}), ~~j=1,2, ~~y(P)=-\frac{1}{2}\sqrt{\mathscr{R}_{4n+4}(z)}, ~~\textrm{for}~~ P\in \Pi_-, \\
&\lim_{P\rightarrow P_0}\psi_j(P,x,t_{p})= \lim_{P\rightarrow P_0}\psi_j(P^*,x,t_{p}),~~\textrm{for}~~j=1,2,~~P\in \Pi_\pm,~~P_0\in\mathcal{C},
\end{align*}
where we choose the branches of $\sqrt{\mathscr{R}_{2n+2}(\cdot)}$ satisfying
$$\lim_{z\rightarrow\infty}\frac{\mathscr{G}_{n+1}}{\sqrt{\mathscr{R}_{2n+2}(z)}}=1.$$
Moreover,
let
\begin{align*}
  \hat{\mu}_j(x,y,t_p)=&\Big(\mu_j(x,y,t_{p}), -2\sqrt{\mathscr{R}_{4n+4}(\mu_j(x,y,t_{p}))}\Big)\in \Pi_-,\\
   \hat{\nu}_j(x,y,t_p)=&\Big(\nu_j(x,y,t_{p}), 2\sqrt{\mathscr{R}_{4n+4}(\mu_j(x,y,t_{p}))}\Big)\in \Pi_+,~~j=1,\ldots, n.
\end{align*}
Based on above preparations, we shall study explicit forms of Baker-Akhiezer function $\psi_j(P),j=1,2$.
\newtheorem{thm4.2}[lem4.1]{Theorem}
\begin{thm4.2}
Suppose $q,r\in C^{\infty}(\mathbb{R}^{2+1})$.
Then Baker-Akhiezer function satisfying condition (\ref{4.1k}) can be expressed as
 \begin{align}
 \psi_1(P,x,y,t_{p})=&
 ~\sqrt{\frac{\mathscr{F}_{2n+1}(\sqrt{\eta},x,y,t_{p})}{\mathscr{F}_{2n+1}(\sqrt{\eta},x_0,y_0,t_{p,0})}}
 \exp\Big(- 2\int_{x_0}^x\frac{q(x^\prime,y,t_{p})\sqrt{\eta}y(P)}{\mathscr{F}_{2n+1}(\sqrt{\eta},x^\prime,y,t_{p})}dx^\prime \nonumber \\
&-  2\int_{y_0}^y\frac{y(P)\widehat{F}_3(\sqrt{\eta},x_0,y^\prime,t_{p})}{\mathscr{F}_{2n+1}(\sqrt{\eta},x_0,y^\prime,t_{p})}dy^\prime
-
2\int_{t_{p,0}}^{t_{p}}\frac{y(P)\widehat{F}_{2p+1}(\sqrt{\eta},x_0,y_0,t^\prime)}{\mathscr{F}_{2n+1}(\sqrt{\eta},x_0,y_0,t^\prime)}dt^\prime\Big),
\label{4.10kp}\\
\psi_2(P,x,y,t_{p})=&~\sqrt{\frac{\mathscr{H}_{2n+1}(\sqrt{\eta},x,y,t_{p})}{\mathscr{F}_{2n+1}(\sqrt{\eta},x_0,y_0,t_{p,0})}}
\exp\Big(2 \int_{x_0}^x\frac{r(x^\prime,y,t_{p})\sqrt{\eta}y(P)}{\mathscr{F}_{2n+1}(\sqrt{\eta},x^\prime,y,t_p)}dx^\prime \nonumber \\
&-2\int_{y_0}^y\frac{y(P)\widehat{H}_3(\sqrt{\eta},x_0,y^\prime,t_p)}{\mathscr{F}_{2n+1}(\sqrt{\eta},x_0,y^\prime,t_p)}dy^\prime
-2\int_{t_{p,0}}^{t_{p}}
\frac{y(P)\widehat{H}_{2p+1}(\sqrt{\eta},x_0,y_0,t^\prime)}{\mathscr{F}_{2n+1}(\sqrt{\eta},x_0,y_0,t^\prime)}dt^\prime\Big),
\label{4.11kp}
 \end{align}
 where $\eta=z^2$ denotes the projection of $P$ to complex sphere.
\end{thm4.2}
\proof
By (\ref{2.14k}), (\ref{2.19k})-(\ref{2.24k}), (\ref{4.5k}), one obtains
\begin{align}
  \psi_{1,x}^\pm(z)=&-\left(z^2-\frac{1}{4}qr\right)\psi_1^\pm(z)+qz\psi_2^\pm(z)\nonumber\\
  =&\left[\left(-\frac{qz\mathscr{G}_{2n+2}(z)}{\mathscr{F}_{2n+1}(z)}
  +\frac{1}{2}\frac{\mathscr{F}_{2n+1,x}(z)}{\mathscr{F}_{2n+1}(z)}\right)
  +qz\frac{\psi_2^\pm(z)}{\psi_1^\pm(z)}\right]\psi_1^\pm(z)\nonumber\\
  =&\left[\left(-\frac{qz\mathscr{G}_{2n+2}(z)}{\mathscr{F}_{2n+1}(z)}
  +\frac{1}{2}\frac{\mathscr{F}_{2n+1,x}(z)}{\mathscr{F}_{2n+1}(z)}\right)
  +q z\frac{\mathscr{H}_{2n+1}(z)}{\mathscr{G}_{2n+2}(z)\pm\sqrt{\mathscr{R}_{4n+4}(z)}}\right]\psi_1^\pm(z)\nonumber\\
  =& \frac{\mp q z \sqrt{\mathscr{R}_{4n+4}(z)}+\frac{1}{2}\mathscr{F}_{2n+1,x}(z)}{\mathscr{F}_{2n+1}(z)}\psi_1^\pm(z),\\
   \psi_{1,t_{j}}^\pm(z)=&-\widehat{G}_{2j+2}(z)\psi_1^\pm(z)+\widehat{F}_{2j+1}(z)\psi_2^\pm(z)\nonumber\\
  =&\left[\left(-\frac{\widehat{F}_{2j+1}(z)\mathscr{G}_{2n+2}(z)}{\mathscr{F}_{2n+1}(z)}
  +\frac{1}{2}\frac{\mathscr{F}_{2n+1,t_{j}}(z)}{\mathscr{F}_{2n+1}(z)}\right)
  +\widehat{F}_{2j+1}(z)\frac{\psi_2^\pm(z)}{\psi_1^\pm(z)}\right]\psi_1^\pm(z)\nonumber\\
  =&\left[\left(-\frac{\widehat{F}_{2j+1}(z)\mathscr{G}_{2n+2}(z)}{\mathscr{F}_{2n+1}(z)}
  +\frac{1}{2}\frac{\mathscr{F}_{2n+1,t_{j}(z)}}{\mathscr{F}_{2n+1}(z)}\right)
  +\widehat{F}_{2j+1}(z) \frac{\mathscr{H}_{2n+1}(z)}{\mathscr{G}_{2n+2}(z)\pm\sqrt{\mathscr{R}_{4n+4}(z)}}\right]\psi_1^\pm(z)\nonumber\\
  =& \frac{\mp \widehat{F}_{2j+1}(z) \sqrt{\mathscr{R}_{4n+4}(z)}+\frac{1}{2}\mathscr{F}_{2n+1,t_{j}}(z)}{\mathscr{F}_{2n+1}(z)}\psi_1^\pm(z),~~j=2,p,
  \end{align}
  and similarly
  \begin{align}
    \psi_{2,x}^\pm(z)=&~rz\psi_1^\pm(z)+\left(z^2-\frac{1}{4}qr\right)\psi_2^\pm(z)\nonumber\\
  =&~\left[rz\frac{\psi_1^\pm(z)}{\psi_2^\pm(z)}
  +\left(\frac{\mathscr{H}_{2n+1,x}(z)}{2\mathscr{H}_{2n+1}(z)}
  -\frac{rz\mathscr{G}_{2n+2}(z)}{\mathscr{H}_{2n+1}(z)}\right)\right]\psi_2^\pm(z)\nonumber\\
  =&~\left[rz \frac{\mathscr{G}_{2n+2}(z)\pm \sqrt{\mathscr{R}_{4n+4}(z)}}{\mathscr{H}_{2n+1}(z)}
  +\left(\frac{\mathscr{H}_{2n+1,x}(z)}{2\mathscr{H}_{2n+1}(z)}
  -\frac{rz\mathscr{G}_{2n+2}(z)}{\mathscr{H}_{2n+1}(z)}\right)\right]\psi_2^\pm(z)\nonumber\\
  =&~\frac{\pm  rz\sqrt{\mathscr{R}_{4n+4}(z)}+\frac{1}{2}\mathscr{H}_{2n+1,x}(z)}{\mathscr{H}_{2n+1}(z)}\psi_2^\pm(z),\\
  \psi_{2,t_{j}}^\pm(z)=&~-\widehat{H}_{2j+1}(z)\psi_1^\pm(z)+\widehat{G}_{2j+2}(z)\psi_2^\pm(z)\nonumber\\
  =&~\left[-\widehat{H}_{2j+1}(z)\frac{\psi_1^\pm(z)}{\psi_2^\pm(z)}+
  \left(\frac{\mathscr{H}_{2n+1,t_{j}}(z)}{2\mathscr{H}_{2n+1}(z)}+\widehat{H}_{2j+1}(z)
  \frac{\mathscr{G}_{2n+2}(z)}{\mathscr{H}_{2n+1}(z)}\right)\right]\psi_2^\pm(z)\nonumber\\
  =&~\left[-\widehat{H}_{2j+1}(z)\frac{\mathscr{G}_{2n+2}(z)\pm \sqrt{\mathscr{R}_{4n+4}(z)}}{\mathscr{H}_{2n+1}(z)}+\left(\frac{\mathscr{H}_{2n+1,t_{j}}(z)}{2\mathscr{H}_{2n+1}(z)}
  +\widehat{H}_{2j+1}(z)\frac{\mathscr{G}_{2n+2}(z)}{\mathscr{H}_{2n+1}(z)}\right)\right]\psi_2^\pm(z)\nonumber\\
  =&~\frac{\mp \widehat{H}_{2j+1}(z) \sqrt{\mathscr{R}_{4n+4}(z)}+\frac{1}{2}\mathscr{H}_{2n+1,t_{j}}(z)}{\mathscr{H}_{2n+1}(z)}\psi_2^\pm(z),~~j=2,p.
\end{align}
Thus, we have
\begin{align}
  d \ln (\psi_1^\pm(z,x,y,t_{p}))=&~\left(\frac{\mp q z \sqrt{\mathscr{R}_{4n+4}(z)}+\frac{1}{2}\mathscr{F}_{2n+1,x}(z)}{\mathscr{F}_{2n+1}(z)}\right)dx \nonumber\\
   &~+
  \left(\frac{\mp \widehat{F}_3(z) \sqrt{\mathscr{R}_{4n+4}(z)}+\frac{1}{2}\mathscr{F}_{2n+1,y}(z)}{\mathscr{F}_{2n+1}(z)}\right) dy\nonumber\\
  &~+
   \left(\frac{\mp \widehat{F}_{2p+1}(z) \sqrt{\mathscr{R}_{4n+4}(z)}+\frac{1}{2}\mathscr{F}_{2n+1,t_{p}}(z)}{\mathscr{F}_{2n+1}(z)}\right) dt_{p}
\end{align}
and
\begin{align}
  d \ln (\psi_2^\pm(z,x,y,t_{p}))=&~\left(\frac{\pm rz \sqrt{\mathscr{R}_{4n+4}(z)}+\frac{1}{2}\mathscr{H}_{2n+1,x}(z)}{\mathscr{H}_{2n+1}(z)}\right)dx \nonumber\\
   &~+
  \left(\frac{\mp \widehat{H}_3(z) \sqrt{\mathscr{R}_{4n+4}(z)}+\frac{1}{2}\mathscr{H}_{2n+1,y}(z)}{\mathscr{H}_{2n+1}(z)}\right) dy\nonumber\\
  &~+\left(
   \frac{\mp \widehat{H}_{2p+1}(z) \sqrt{\mathscr{R}_{4n+4}(z)}+\frac{1}{2}\mathscr{H}_{2n+1,t_{p}}(z)}{\mathscr{H}_{2n+1}(z)}\right) dt_{p}.
\end{align}
According to relations (\ref{4.6kp})-(\ref{4.8kp}), it follows that
the integrals
\begin{equation}\label{4.18kp}
 \int_{(x_0,y_0,t_{p,0})}^{(x,y,t_{p})}d \ln \big(\frac{\psi_j^\pm(z,x,y,t_{p})}{\sqrt{\mathscr{F}_{2n+1}(z,x,y,t_{p})}}\big),~~j=1,2
\end{equation}
is independent of the path. Therefore taking into account the normalization condition $\psi_1(P,x_0,y_0,t_{p,0})=1$, and
choosing
a special path
$$(x_0,y_0,t_{p,0})\rightarrow (x_0,y_0,t_{p})\rightarrow (x_0,y,t_{p}) \rightarrow (x,y,t_{p})$$
in (\ref{4.18kp}),
we finally obtain (\ref{4.10kp}) and (\ref{4.11kp}).  \qed\vspace{0.3cm}

Basic properties of Baker-Akhiezer
functions $\psi_j(P),j=1,2$ are summarized in the following result.

\newtheorem{lem4.4}[lem4.1]{Lemma}
\begin{lem4.4}
Suppose $q,r\in C^{\infty}(\mathbb{R}^{2+1})$ and $P\in X\backslash\{P_{\infty\pm}\}$.
Then the Baker-Akhiezer function derived in (\ref{4.10kp}), (\ref{4.11kp}) satisfy
\begin{align}
  &\psi_1(P,x,y,t_{p})\psi_1(P^*,x,y,t_{p})=
  \frac{\mathscr{F}_{2n+1}(\sqrt{\eta},x,y,t_{p})}{\mathscr{F}_{2n+1}(\sqrt{\eta},x_0,y_0,t_{p,0})}, \label{4.19kp}\\
  &\psi_2(P,x,y,t_{p})\psi_2(P^*,x,y,t_{p})=
  \frac{\mathscr{H}_{2n+1}(\sqrt{\eta},x,y,t_{p})}{\mathscr{F}_{2n+1}(\sqrt{\eta},x_0,y_0,t_{p,0})}, \label{4.20kp}\\
  & \psi_1(P,x,y,t_{p})\psi_2(P^*,x,y,t_{p})
  =\frac{\mathscr{G}_{2n+2}(\sqrt{\eta},x,y,t_p)
  +\sqrt{\mathscr{R}_{4n+4}(\sqrt{\eta})}}{\mathscr{F}_{2n+1}(\sqrt{\eta},x_0,y_0,t_{p,0})},\label{4.21kp}\\
  &\psi_1(P,x,y,t_{p})\psi_2(P^*,x,y,t_{p})+\psi_1(P^*,x,y,t_{p})\psi_2(P,x,y,t_{p})\nonumber\\
  &~~~~~~~~~~~~~~~~~~~~~~~~~~~~~~~=
  \frac{2\mathscr{G}_{2n+2}(\sqrt{\eta},x,y,t_{p})}{\mathscr{F}_{2n+1}(\sqrt{\eta},x_0,y_0,t_{p,0})},\label{4.22kp}\\
  &\psi_1(P,x,y,t_{p})\psi_2(P^*,x,y,t_{p})-\psi_1(P^*,x,y,t_{p})\psi_2(P,x,y,t_{p})\nonumber\\
  &~~~~~~~~~~~~~~~~~~~~~~~~~~~~~~~=
  \frac{2\sqrt{\mathscr{R}_{4n+4}(\sqrt{\eta})}}{\mathscr{F}_{2n+1}(\sqrt{\eta},x_0,y_0,t_{p,0})}.\label{4.23kp}
\end{align}
\proof It can be easily seen that (\ref{4.19kp}) and (\ref{4.20kp}) hold by (\ref{4.10kp}) and (\ref{4.11kp}), respectively. Then from (\ref{4.19kp}), (\ref{4.20kp}) and the fact that $$\left(\frac{1}{2}(\psi_1(P)\psi_2(P^*)+\psi_1(P^*)\psi_2(P)), \psi_1(P)\psi_1(P^*), \psi_2(P)\psi_2(P^*)\right)$$
 and
 $$\left(\frac{\mathscr{G}_{2n+2}(\sqrt{\eta},x,y,t_{p})}{\mathscr{F}_{2n+1}(\sqrt{\eta},x_0,y_0,t_{p,0})}, \frac{\mathscr{F}_{2n+1}(\sqrt{\eta},x,y,t_{p})}{\mathscr{F}_{2n+1}(\sqrt{\eta},x_0,y_0,t_{p,0})}, \frac{\mathscr{H}_{2n+1}(\sqrt{\eta},x,y,t_{p})}{\mathscr{F}_{2n+1}(\sqrt{\eta},x_0,y_0,t_{p,0})}\right)$$  are both solutions of
 linear system (\ref{2.16k})-(\ref{2.24k}) satisfying the same initial condition, one gets (\ref{4.22kp}). Using (\ref{4.19kp}), (\ref{4.20kp}) and (\ref{4.22kp}), we have
\begin{align}
  &\psi_1(P,x,y,t_{p})\psi_2(P^*,x,y,t_{p})-\psi_1(P^*,x,y,t_{p})\psi_2(P,x,y,t_{p})\nonumber\\
 =& \sqrt{(\psi_1(P)\psi_2(P^*)+\psi_1(P^*)\psi_2(P))^2-4\psi_1(P)\psi_1(P^*)\psi_2(P)\psi_2(P^*)}\nonumber\\
 =&\frac{2\sqrt{\mathscr{R}_{4n+4}(\sqrt{\eta})}}{\mathscr{F}_{2n+1}(\sqrt{\eta},x_0,y_0,t_{p,0})}.
\end{align}
Finally, (\ref{4.21kp}) is the direct result of (\ref{4.22kp}), (\ref{4.23kp}).
\qed

\end{lem4.4}

 Next we shall study the analytic property and asymptotic behavior of $\psi_1(P)\psi_2(P^*)$, and
 $\psi_j(P), j=1,2.$

\newtheorem{lem4.5a}[lem4.1]{Lemma}
\begin{lem4.5a}
The function $\psi_1(P)\psi_2(P^*)$ is a meromorphic function on $X$ with divisor
\begin{equation}\label{4.25dp}
(\psi_1(P)\psi_2(P^*))=\mathcal{D}_{\underline{\hat{\nu}}^*(x,y,t_p)\underline{\hat{\mu}}(x,y,t_p)P_{\infty+}}-\mathcal{D}_{\underline{\hat{\mu}}^*(x_0,y_0,t_0)\underline{\hat{\mu}}(x_0,y_0,t_0)P_{\infty-}},
\end{equation}
where we abbreviate $\underline{\hat{\mu}}=(\hat{\mu}_1,\ldots,\hat{\mu}_n), \underline{\hat{\nu}}=(\hat{\nu}_1,\ldots,\hat{\nu}_n).$
Moreover, using the local coordinate $\zeta=z^{-1}$ near $P_{\infty\pm},$ we have
\begin{align}\label{4.26dp}
   \psi_1(P,x,y,t_p)\psi_2(P^*,x,y,t_p)\overset{\zeta\rightarrow 0}{=}
   \begin{cases}
   \frac{2}{q(x_0,y_0,t_{p,0})}\zeta^{-1}+O(1), & \textrm{as}~~P\rightarrow P_{\infty+}, \cr
   - \frac{ q(x,y,t_{p})r(x,y,t_{p})}{ 2q(x_0,y_0,t_{p,0})}\zeta+O(\zeta^2), & \textrm{as}~~P\rightarrow P_{\infty-}. \cr
   \end{cases}
\end{align}
\end{lem4.5a}
\proof
Noticing  (\ref{3.3zp}), (\ref{2.22}), (\ref{4.21kp}) and
\begin{align*}
   \psi_1(P,x,y,t_{p})\psi_2(P^*,x,y,t_{p})=
   &~\frac{\mathscr{G}_{2n+2}(\sqrt{\eta},x,y,t_{p})
   +\sqrt{\mathscr{R}_{4n+4}(\sqrt{\eta})}}{\mathscr{F}_{2n+1}(\sqrt{\eta},x_0,y_0,t_{p,0})}\nonumber\\
   =&~\frac{\mathscr{G}_{2n+2}(\sqrt{\eta},x,y,t_{p})
   +\sqrt{\mathscr{R}_{4n+4}(\sqrt{\eta})}}{\mathscr{F}_{2n+1}(\sqrt{\eta},x,y,t_{p})}\frac{\mathscr{F}_{2n+1}(\sqrt{\eta},x,y,t_{p})}{\mathscr{F}_{2n+1}(\sqrt{\eta},x_0,y_0,t_{p,0})}\nonumber\\
   =&~\frac{\mathscr{H}_{2n+1}(\sqrt{\eta},x,y,t_{p})}{\mathscr{G}_{2n+2}(\sqrt{\eta},x,y,t_{p})
   -\sqrt{\mathscr{R}_{4n+4}(\sqrt{\eta})}}\frac{\mathscr{F}_{2n+1}(\sqrt{\eta},x,y,t_{p})}{\mathscr{F}_{2n+1}(\sqrt{\eta},x_0,y_0,t_{p,0})},\nonumber\\
\end{align*}
 one easily proves (\ref{4.25dp}) and (\ref{4.26dp}). \qed \vspace{0.3cm}

Now we turn to study the analytic structure of $\psi_j(P,x,y,t_{p})$ on $X\backslash\{P_{\infty\pm}\}.$

\newtheorem{thm4.5}[lem4.1]{Theorem}
\begin{thm4.5}
 Assume auxiliary spectrum points $\mu_j(x,y,t_{p})$, $\nu_j(x,y,t_{p}),$ \linebreak $j=1,\ldots,n,$ are mutually distinct and finite for all $(x,y,t_{p})\in\Omega,$
where $\Omega\in \mathbb{R}^3$ is an open interval.
Moreover, let $P\in X\backslash\{P_{\infty\pm}\}.$
Then
\begin{itemize}
\item[\textbf{I.}]
$\psi_1(P,x,y,t_{p})$ and $\psi_2(P,x,y,t_{p})$ are meromorphic on $P\in X\backslash\{P_{\infty\pm}\}.$ Their divisor of poles coincides with $\mathcal{D}_{\underline{\hat{\mu}}(x_0,y_0,t_{p,0})}.$

\item[\textbf{II.}] The divisor of zeros for $\psi_1(P,x,y,t_{p})$ and $\psi_2(P,x,y,t_{p})$
coincides with $\mathcal{D}_{\underline{\hat{\mu}}(x,y,t_{r+1})}$ and $\mathcal{D}_{\underline{\hat{\nu}}(x,y,t_{p})},$ respectively.

\item[\textbf{III.}] As $P\rightarrow P_{\infty\pm},$ the asymptotic behavior of $(\psi_1(P,x,y,t_{p}),\psi_2(P,x,y,t_{p}))^T$ is given by the equations,
    \begin{align}
     \left(
       \begin{array}{c}
         \psi_1(P,x,y,t_{p}) \\
         \psi_2(P,x,y,t_{p}) \\
       \end{array}
     \right)=&~\left(
                     \begin{array}{c}
                       \frac{q(x,y,t_{p})}{q(x_0,y_0,t_{p,0})}+ O(\zeta) \\                       -\frac{q(x,y,t_p)r(x,y,t_p)}{2q(x_0,y_0,t_{p,0})}\zeta+O(\zeta^2) \\
                     \end{array}
                   \right)\times\exp\Big(-(x-x_0)\zeta^{-1}
     \nonumber\\
     &-(y-y_0)\zeta^{-2}
     -(t_{p}-t_{p,0})\zeta^{-(p+1)}\nonumber\\
     &+\Big(-\frac{1}{4}\int_{x_0}^x
     q(x^\prime,y,t_p)r(x^\prime,y,t_p)dx^\prime+\frac{1}{2}\int_{y_0}^yg_{4}(x_0,y^\prime,t_p)dy^\prime
       \nonumber\\
     &+\frac{1}{2}\int_{t_{p,0}}^{t_p}g_{2p+2}(x_0,y_0,t^\prime)dt^{\prime}\Big)\Big),~~\textrm{at}~~P\rightarrow P_{\infty+},\label{4.27dp}
    \end{align}
and
\begin{align}\label{4.28dp}
     \left(
       \begin{array}{c}
         \psi_1(P,x,y,t_{p}) \\
         \psi_2(P,x,y,t_{p}) \\
       \end{array}
     \right)=&~\left(
                     \begin{array}{c}
                       1+O(\zeta) \\
                       \frac{2}{q(x,y,t_p)}\zeta^{-1}+O(1) \\
                     \end{array}
                   \right)\times\exp\Big((x-x_0)\zeta^{-1}
     \nonumber\\
                   &+(y-y_0)\zeta^{-2}+(t-t_{p,0})\zeta^{-(p+1)}\nonumber\\
     &+\Big(\frac{1}{4}\int_{x_0}^xq(x^\prime,y,t_p)r(x^\prime,y,t_p)dx^\prime-\frac{1}{2}\int_{y_0}^yg_{4}(x_0,y^\prime,t_p)dy^\prime
       \nonumber\\
     &-\frac{1}{2}\int_{t_{p,0}}^{t_p}g_{2p+2}(x_0,y_0,t^\prime)dt^{\prime}\Big)\Big),~~\textrm{at}~~P\rightarrow P_{\infty-}.
    \end{align}
\end{itemize}
\end{thm4.5}

\proof First we study the function $\psi_1$. By (\ref{2.28l}), (\ref{2.29m}), (\ref{2.29n}) it follows
\begin{align}
 - 2\frac{q(x^\prime,y,t_{p})\sqrt{\eta}y(P)}{\mathscr{F}_{2n+1}(\sqrt{\eta},x^\prime,y,t_{p})}
 &\overset{P\rightarrow \hat{\mu}_j(x^\prime,y,t_{p})}{=} \partial_{x^\prime} \ln \sqrt{\eta-\mu_j(x^\prime,y,t_{p})},\\
  -\frac{2y(P)\widehat{F}_3(\sqrt{\eta},x_0,y^\prime,t_{p})}{\mathscr{F}_{2n+1}(\eta,x_0,y^\prime,t_{p})}&\overset{P\rightarrow \hat{\mu}_j(x_0,y^\prime,t_{p})}{=} \partial_{y^\prime} \ln \sqrt{\eta-\mu_j(x_0,y^\prime,t_{p})},\\
  -\frac{2y(P)\widehat{F}_{2p+1}(\sqrt{\eta},x_0,y_0,t^\prime)}{\mathscr{F}_{2n+1}(\sqrt{\eta},x_0,y_0,t^\prime)}&\overset{P\rightarrow \hat{\mu}_j(x_0,y_0,t^\prime)}{=}~~ \partial_{t^\prime} \ln \sqrt{\eta-\mu_j(x_0,y_0,t^\prime)},
\end{align}
and hence one obtains
\begin{align}
 & \exp\Big(-2\int_{x_0}^x\frac{q(x^\prime,y,t_{p})\sqrt{\eta}y(P)}{\mathscr{F}_{2n+1}(\sqrt{\eta},x^\prime,y,t_{p})}dx^\prime- 2\int_{y_0}^y\frac{y(P)\widehat{F}_3(\sqrt{\eta},x_0,y^\prime,t_{p})}{\mathscr{F}_{2n+1}(\sqrt{\eta},x_0,y^\prime,t_{p})}dy^\prime\nonumber\\
 &-2\int_{t_{p,0}}^{t_{p}}\frac{y(P)\widehat{F}_{2p+1}(\sqrt{\eta},x_0,y_0,t^\prime)}{\mathscr{F}_{2n+1}(\sqrt{\eta},x_0,y_0,t^\prime)}dt^\prime\Big)\nonumber\\
 =&\exp\Big(\int_{x_0}^x\partial_{x^\prime} \ln \sqrt{\eta-\mu_j(x^\prime,y,t_{p})}dx^\prime + \int_{y_0}^y
 \partial_{y^\prime} \ln \sqrt{\eta-\mu_j(x_0,y^\prime,t_{p})}dy^\prime\nonumber\\
 &+ \int_{t_{p,0}}^{t_{p}} \partial_{t^\prime} \ln \sqrt{\eta-\mu_j(x_0,y_0,t^\prime)}dt^\prime+O(1)
  \Big)\nonumber\\
 =&\begin{cases}
 \sqrt{\eta-\mu_j(x_0,y_0,t_{p,0})}^{-1},& P\rightarrow \hat{\mu}_j(x_0,y_0,t_{p,0}),\cr
 \sqrt{\eta-\mu_j(x,y,t_{p})},& P\rightarrow \hat{\mu}_j(x,y,t_{p})\neq \hat{\mu}_j(x_0,y_0,t_{p,0}),\cr
  O(1),& P\rightarrow \hat{\mu}_j(x,y,t_{p})= \hat{\mu}_j(x_0,y_0,t_{p,0}),\cr
  O(1),& P\rightarrow \textrm{other points}\neq  \hat{\mu}_j(x,y,t_{p}),~ \hat{\mu}_j(x_0,y_0,t_{p,0}).
 \end{cases}
 \end{align}
Then taking into account (\ref{4.10kp}), one proves {\bf I.} and {\bf II.} for $\psi_1$. Next we study the asymptotic behaviour of $\psi_1$ near $P_{\infty\pm}.$ Again using (\ref{4.10kp}) and local coordinate $\eta={z}^{-1}$ near $P_{\infty\pm}$,
one infers
\begin{align}
\psi_1(P,x,y,t_{p})=&~\sqrt{\frac{\mathscr{F}_{2n+1}(\sqrt{\eta},x,y,t_{p})}{\mathscr{F}_{2n+1}(\sqrt{\eta},x_0,y_0,t_{p,0})}}
\exp\Big(-2\int_{x_0}^x\frac{q(x^\prime,y,t_{p})y(P)}{\mathscr{F}_{2n+1}(\sqrt{\eta},x^\prime,y,t_{p})}dx^\prime \nonumber \\
&-2\int_{y_0}^y\frac{y(P)\widehat{F}_3(\sqrt{\eta},x_0,y^\prime,t_{p})}{\mathscr{F}_{2n+1}(\sqrt{\eta},x_0,y^\prime,t_{p})}dy^\prime
-2\int_{t_{p,0}}^{t_{p}}\frac{y(P)\widehat{F}_{2p+1}(z,x_0,y_0,t^\prime)}{\mathscr{F}_{2n+1}(\sqrt{\eta},x_0,y_0,t^\prime)}dt^\prime\Big)\nonumber\\
=&\exp\Big(\int_{x_0}^x\left(-2\frac{q(x^\prime,y,t_{p})y(P)}{\mathscr{F}_{2n+1}(\sqrt{\eta},x^\prime,y,t_{p})}
+\frac{\mathscr{F}_{2n+1,x^\prime}(\sqrt{\eta},x^\prime,y,t_{p})}{2\mathscr{F}_{2n+1}(\sqrt{\eta},x^\prime,y,t_{p})}\right)dx^\prime \nonumber \\
&+\int_{y_0}^y\left(-
2\frac{y(P)\widehat{F}_3(\sqrt{\eta},x_0,y^\prime,t_{p})}{\mathscr{F}_{2n+1}(\sqrt{\eta},x_0,y^\prime,t_{p})}+ \frac{\mathscr{F}_{2n+1,y^\prime}(\sqrt{\eta},x_0,y^\prime,t_{p})}{2\mathscr{F}_{2n+1}(\sqrt{\eta},x_0,y^\prime,t_{p})}
\right)dy^\prime \nonumber\\
&+\int_{t_{p,0}}^{t_{p}}\left(-2\frac{y(P)\widehat{F}_{2p+1}(\sqrt{\eta},x_0,y_0,t^\prime)}{\mathscr{F}_{2n+1}(\sqrt{\eta},x_0,y_0,t^\prime)}
+\frac{\mathscr{F}_{2n+1,t^\prime}(\sqrt{\eta},x_0,y_0,t^\prime)}{2\mathscr{F}_{2n+1}(\sqrt{\eta},x_0,y_0,t^\prime)}\right)dt^\prime\Big)\nonumber\\
=&\begin{cases}
\exp\Big(\int_{x_0}^x \left( \frac{-2q(x^\prime,y,t_{p})}{\sum_{j=0}^\infty\hat{f}_{2j+1}(x^\prime,y,t_{p})\zeta^{j+1}}
+\frac{q_{x^\prime}(x^\prime,y,t_{p})}{2q(x^\prime,y,t_{p})}\right)dx^\prime\nonumber\\
~~~~+\int_{y_0}^y \left( \frac{-2\widehat{F}_3(x_0,y^\prime,t_{p})}{\sum_{j=0}^\infty\hat{f}_{2j+1}(x_0,y^\prime,t_{p})\zeta^{j+1}}
+\frac{q_{y^\prime}(x_0,y^\prime,t_{p})}{2q(x_0,y^\prime,t_{p})}\right)dy^\prime\nonumber\\
~~~~+\int_{t_{p,0}}^{t_{p}} \left( \frac{-2\widehat{F}_{2p+1}(x_0,y_0,t^\prime)}{\sum_{j=0}^\infty\hat{f}_{2j+1}(x_0,y_0,t^\prime)\zeta^{j+1}}
+\frac{q_{t^\prime}(x_0,y_0,t^\prime)}{2q(x_0,y_0,t^\prime)}\right)dt^\prime\Big),& \textrm{as}~~P\rightarrow P_{\infty-},
\cr
\exp\Big(\int_{x_0}^x \left( \frac{2q(x^\prime,y,t_{p})}{\sum_{j=0}^\infty\hat{f}_{2j+1}(x^\prime,y,t_{p})\zeta^{j+1}}
+\frac{q_{x^\prime}(x^\prime,y,t_{p})}{2q(x^\prime,y,t_{p})}\right)dx^\prime\nonumber\\
~~~~+\int_{y_0}^y \left( \frac{2\widehat{F}_3(x_0,y^\prime,t_{p})}{\sum_{j=0}^\infty\hat{f}_{2j+1}(x_0,y^\prime,t_{p})\zeta^{j+1}}
+\frac{q_{y^\prime}(x_0,y^\prime,t_{p})}{2q(x_0,y^\prime,t_{p})}\right)dy^\prime\nonumber\\
~~~~+\int_{t_{p,0}}^{t_{p}} \left( \frac{2\widehat{F}_{2p+1}(x_0,y_0,t^\prime)}{\sum_{j=0}^\infty\hat{f}_{2j+1}(x_0,y_0,t^\prime)\zeta^{j+1}}
+\frac{q_{t^\prime}(x_0,y_0,t^\prime)}{2q(x_0,y_0,t^\prime)}\right)dt^\prime\Big),&\textrm{as}~~P\rightarrow P_{\infty+},
\end{cases}\nonumber\\
=&\begin{cases}
\frac{q(x,y,t_{p})}{q(x_0,y_0,t_{p,0})}\exp\Big(-(x-x_0)\zeta^{-1}
 -(y-y_0)\zeta^{-2}-(t_p-t_{p,0})\zeta^{-(p+1)}\nonumber\\+(-\frac{1}{4}\int_{x_0}q(x^\prime,y,t_p)r(x^\prime,y,t_p)dx^\prime+\frac{1}{2}\int_{y_0}^yg_{4}(x_0,y^\prime,t_p)dy^\prime
 \nonumber\\
 +\frac{1}{2}\int_{t_{p,0}}^{t_p}g_{2p+2}(x_0,y_0,t^\prime)dt^{\prime})+O(\zeta)\Big), ~~~~\textrm{as}~~P\rightarrow P_{\infty-},
\cr
\exp\Big((x-x_0)\zeta^{-1}
 +(y-y_0)\zeta^{-2}+(t_p-t_{p,0})\zeta^{-(p+1)}\nonumber\\
 +(\frac{1}{4}\int_{x_0}q(x^\prime,y,t_p)r(x^\prime,y,t_p)dx^\prime-\frac{1}{2}\int_{y_0}^yg_{4}(x_0,y^\prime,t_p)dy^\prime
 \nonumber\\
 -\frac{1}{2}\int_{t_{p,0}}^{t_p}g_{2p+2}(x_0,y_0,t^\prime)dt^{\prime})+O(\zeta)\Big), ~~~~\textrm{as}~~P\rightarrow P_{\infty+}.
\end{cases}\\\label{4.33dp}
\end{align}
Here we have used asymptotic spectral expansion
\begin{equation}\label{4.35mp}
  \frac{\mathscr{F}_{2n+1}(\sqrt{\eta},x,y,t_{p})}{y(P)}=\pm \sum_{j=0}^\infty\hat{f}_{2j+1}(x,y,t_{p})\zeta^{j+\frac{3}{2}},~~\textrm{as}~~P\rightarrow P_{\infty\pm},
\end{equation}
in the last equality of (\ref{4.33dp}),
and the expression (\ref{4.35mp}) can be obtained by induction.
 Finally,
the statements for $\psi_2$ follows by Lemma 4.5
and (\ref{4.33dp}).
\qed

\section{Algebro-Geometric Solutions}

In this section, we shall study
the theta function representation of Baker-Akhiezer functions $\psi_1(P), \psi_2(P)$ and algebro-geometric solutions of the whole mKP hierarachy. For more details about Riemann surfaces and theta functions we recommend the reference \cite{Farkas}.

\newtheorem{thm5.1}{Theorem}[section]

First, choosing a convenient base point $Q_0 \in
     X \setminus \{P_{\infty\pm}\}$, the Abel maps
      $\underline{A}_{Q_0}(\cdot) $ and
      $\underline{\alpha}_{Q_0}(\cdot)$ are defined by
         \begin{eqnarray*}
           \underline{A}_{Q_0}:X \rightarrow
           J(X)&=&\mathbb{C}^{n}/L_{n},
         \end{eqnarray*}
         \begin{eqnarray*}
          P \mapsto \underline{A}_{Q_0} (P)&=& (A_{Q_0,1}(P),\ldots,
           A_{Q_0,n} (P)) \\
           &=&\left(\int_{Q_0}^P\omega_1,\ldots,\int_{Q_0}^P\omega_{n}\right)
           (\mathrm{mod}~L_{n}),
         \end{eqnarray*}
     and
         \begin{eqnarray*}
          && \underline{\alpha}_{Q_0}:
          \mathrm{Div}(X) \rightarrow
          J(X),\\
          &&~~~~~\qquad \mathcal{D} \mapsto \underline{\alpha}_{Q_0}
          (\mathcal{D})= \sum_{P\in \mathcal{K}_{n}}
           \mathcal{D}(P)\underline{A}_{Q_0} (P),
         \end{eqnarray*}
    where $L_{n}=\{\underline{z}\in \mathbb{C}^{n}|
           ~\underline{z}=\underline{N}+\Gamma\underline{M},
           ~\underline{N},~\underline{M}\in \mathbb{Z}^{n}\}$, and
  $\Gamma$, $\underline{\Xi}_{Q_0}$ are the Riemann matrix and the vector of Riemann
     constants, respectively. Moreover,
      we choose a homology basis $\{a_{j},b_{j}\}_{j=1}^{n}$ on $X$ in such a way that the intersection matrix of the cycles satisfies
\begin{equation}
a_{j}\circ b_{k}=\delta_{j,k}, \quad a_{j}\circ a_{k}=0,\quad b_{j}\circ b_{k}=0,\quad j,~k=1,\ldots,n.
\end{equation}
     For brevity, define the function
      $\underline{z}: X \times
      \sigma^{n} X \rightarrow \mathbb{C}^{n}$ by\footnote{$\sigma^{n} X$=
      $\underbrace{ X\times\ldots\times X}_{n}.$}
     \begin{eqnarray}\label{4.4}
           \underline{z}(P,\underline{Q})&=~\underline{\Xi}_{Q_0}
           -\underline{A}_{Q_0}(P)+\underline{\alpha}_{Q_0}
             (\mathcal{D}_{\underline{Q}}), \nonumber \\
           P\in \mathcal{K}_{n},\,~
           \underline{Q}&=~(Q_1,\ldots,Q_{n})\in
           \sigma^{n}\mathcal{K}_{n},
         \end{eqnarray}
     here $\underline{z}(\cdot,\underline{Q}) $ is
     independent of the choice of base point $Q_0$.
     The Riemann theta
     function $\theta(\underline{z})$ associated with $X$ and the homology is
      defined by
     $$\theta(\underline{z})=\sum_{\underline{n}\in\mathbb{Z}}\exp\left(2\pi i<\underline{n},\underline{z}>+\pi i<\underline{n},\underline{n}\Gamma>\right),\quad \underline{z}\in\mathbb{C}^{n},$$
     where $<\underline{B},\underline{C}>=\overline{\underline{B}}
     \cdot\underline{C}^t=\sum_{j=1}^{N}\overline{B}_jC_j$
     denotes the scalar product in $\mathbb{C}^{n-1}$.

Let $\omega_{P_{\infty\pm},q}^{(2)}$ be the normalized differentials
of the second kind with a unique pole at $P_{\infty\pm}$, respectively,
and principal parts
 \begin{equation}\label{5.13}
 \omega_{P_{\infty\pm},q}^{(2)}\underset{\zeta\rightarrow0}{=}\left(\zeta^{-2-q}+O(1)
\right)d\zeta,~~P\rightarrow P_{\infty\pm},~~\zeta=z^{-1},~~q\in\mathbb{N}_{0}
\end{equation}
with vanishing $a$-periods,
\begin{equation*}
\int_{a_{j}}\omega_{P_{\infty\pm},q}^{(2)}=0,~~ j=1,\ldots,n,
\end{equation*}
and
$\omega_{P_{\infty+}P_{\infty-}}$ be normalized differential of third kind
satisfying
\begin{equation*}
\begin{split}
   \int_{Q_0}^P\omega_{P_{\infty-}P_{\infty+}}=\zeta^{-1}+c_{\infty+}+O(\zeta),~~ P\rightarrow P_{\infty+}, \zeta=z^{-1}, c_{\infty+}\in\mathbb{C},
 \end{split}
 \end{equation*}
 and
 \begin{equation*}
 \begin{split}
 \int_{a_{j}}\omega_{P_{\infty+}P_{\infty-}}=0~~ j=1,\ldots,n.
  \end{split}
\end{equation*}
Moreover, we introduce the notation
 \begin{equation}\label{5.14}
\Omega_{r}^{(2)}=
 \frac{1}{2}(r+1)(\omega_{P_{\infty+},r}^{(2)}-\omega_{P_{\infty-},r}^{(2)}),~~r\in\mathbb{N}_0.
\end{equation}
and
\begin{equation}
 \omega_{r}^{(2)} =\lim_{P\rightarrow P_{\infty+}}\int_{Q_0}^P\Omega_{r}^{(2)}.
\end{equation}

 Next we study the theta function representation for $\psi_1(P)\psi_2(P^*)$.

\newtheorem{thm5.2}[thm5.1]{Lemma}
\begin{thm5.2}
Assume spectral curve defined in (\ref{2.22}) is nonsingular and $q,r$ satisfy (\ref{st2.51}).
Moreover, let $P\in X\backslash\{P_{\infty\pm}\}$ and suppose the divisors $\mathcal{D}_{\underline{\hat{\nu}}(x,y,t_{p})}$, $\mathcal{D}_{\underline{\hat{\mu}}(x,y,t_{p})}$ are nonspecial. Then  \\
\begin{align}\label{5.16a}
&\psi_1(P,x,y,t_{p})\psi_2(P^*,x,y,t_{p})\nonumber\\
=&\frac{2}{q(x_0,y_0,t_{p,0})}\frac{\theta(\underline{z}(P_{\infty+}, \underline{\hat{\mu}}(x_0,y_0,t_{p,0})))\theta(\underline{z}(P_{\infty-}, \underline{\hat{\mu}}(x_0,y_0,t_{p,0})))}{\theta(\underline{z}(P_{\infty+}, \underline{\hat{\nu}}(x,y,t_{p})))\theta(\underline{z}(P_{\infty-}, \underline{\hat{\mu}}(x,y,t_{p})))}\nonumber\\
&\times \frac{\theta(\underline{z}(P^*, \underline{\hat{\nu}}(x,y,t_{p})))\theta(\underline{z}(P, \underline{\hat{\mu}}(x,y,t_{p})))}{\theta(\underline{z}(P^*, \underline{\hat{\mu}}(x_0,y_0,t_{p,0})))\theta(\underline{z}(P,\underline{\hat{\mu}}(x_0,y_0,t_{p,0})))}\nonumber
\\
&\times\exp\Big(\int_{Q_0}^P\omega_{P_{\infty-}P_{\infty+}}-c_{\infty+}\Big).
\end{align}

\end{thm5.2}

\proof 
Let
\begin{align}
 \circledR(P,x,t_{p})=&~\frac{2}{q(x_0,y_0,t_{r+1,0})}\frac{\theta(\underline{z}(P_{\infty+}, \underline{\hat{\mu}}(x_0,y_0,t_{p,0})))\theta(\underline{z}(P_{\infty-}, \underline{\hat{\mu}}(x_0,y_0,t_{p,0})))}{\theta(\underline{z}(P_{\infty+}, \underline{\hat{\nu}}(x,y,t_{p})))\theta(\underline{z}(P_{\infty-}, \underline{\hat{\mu}}(x,y,t_{p})))}\nonumber\\
&\times \frac{\theta(\underline{z}(P^*, \underline{\hat{\nu}}(x,y,t_{p})))\theta(\underline{z}(P, \underline{\hat{\mu}}(x,y,t_{p})))}{\theta(\underline{z}(P^*, \underline{\hat{\mu}}(x_0,y_0,t_{p,0})))\theta(\underline{z}(P, \underline{\hat{\mu}}(x_0,y_0,t_{p,0})))}\nonumber
\\
&\times\exp\Big(\int_{Q_0}^P\omega_{P_{\infty-}P_{\infty+}}-c_{\infty+}\Big).
\end{align}
Then it is not difficult to know divisors of $\psi_1(P,x,y,t_{p})\psi_2(P^*,x,y,t_{p})$ coincides with that of $\circledR(P,x,t_{p})$ on $X$. Therefore, the function $\frac{\psi_1(P,x,y,t_{p})\psi_2(P^*,x,y,t_{p})}{\circledR(P,x,y,t_{p})}$ is a constant independent of $P$ and we denote it by $C(x,y,t_{p})$. Taking $P\rightarrow P_{\infty+}$ and using Lemma 4.5,
one finally obtains
$$C(x,y,t_{p})=\lim_{P\rightarrow P_{\infty+}}\frac{\psi_1(P,x,y,t_{p})\psi_2(P^*,x,y,t_{p})}{\circledR(P,x,y,t_{p})}=1,~~$$
and the relation
$$\psi_1(P,x,y,t_{p})\psi_2(P^*,x,y,t_{p})=\circledR(P,x,y,t_{p})$$ for any $P\in X.$
\qed
\\

Given these preparations, one finally derives the following theta function representation
for Baker-Akhiezer functions $\psi_j(P,x,y,t_{p}),j=1,2,$ and algebro-geometric solutions $q(x,y,t_{p}),r(x,y,t_{p}).$\\

\newtheorem{thm5.3}[thm5.1]{Theorem}
\begin{thm5.3}
Assume spectral curve defined in (\ref{2.22}) is nonsingular and $q,p$ satisfy (\ref{st2.51}).
Moreover, let $P\in X\backslash\{P_{\infty\pm}\}$ and suppose the divisors $\mathcal{D}_{\underline{\hat{\nu}}(x,y,t_{p})}$, $\mathcal{D}_{\underline{\hat{\mu}}(x,y,t_{p})}$ are nonspecial.
Then functions $\psi_1,\psi_2,q,r$ have
the following theta function representations
\begin{align}
  \psi_1(P,x,y,t_{p})=&~C_1(x,y,t_{p})\frac{\theta(\underline{z}(P, \underline{\hat{\mu}}(x,y,t_{p})))}{\theta(\underline{z}(P, \underline{\hat{\mu}}(x_0,y_0,t_{p,0})))}\exp\Big(-(x-x_0)\int_{Q_0}^P\Omega^{(2)}_0 \nonumber\\
  &-(y-y_0)\int_{Q_0}^P\Omega^{(2)}_1-(t_p-t_{p,0})\int_{Q_0}^P\Omega^{(2)}_p\Big),\label{5.15cp}\\
   \psi_2(P,x,y,t_{p})=&~C_2(x,y,t_{p})\frac{\theta(\underline{z}(P, \underline{\hat{\nu}}(x,y,t_{p})))}{\theta(\underline{z}(P, \underline{\hat{\mu}}(x_0,y_0,t_{p,0})))}\exp\Big(-(x-x_0)\int_{Q_0}^P\Omega^{(2)}_0 \nonumber\\
   &-(y-y_0)\int_{Q_0}^P\Omega^{(2)}_1
   -(t_p-t_{p,0})\int_{Q_0}^P\Omega^{(2)}_p+ \sqrt{\int_{Q_0}^P\omega_{P_{\infty+} P_{\infty-}}}\Big),\label{5.16cp}
\end{align}
and
\begin{align}
   q(x,y,t_{p})
 =&~q(x_0,y_0,t_{p,0}) \frac{\theta(\underline{z}(P_{\infty+}, \underline{\hat{\mu}}(x_0,y_0,t_{p,0})))}{\theta(\underline{z}(P_{\infty+}, \underline{\hat{\mu}}(x,y,t_{p})))}
 \frac{\theta(\underline{z}(P_{\infty-}, \underline{\hat{\mu}}(x,y,t_{p})))}{\theta(\underline{z}(P_{\infty-}, \underline{\hat{\mu}}(x_0,y_0,t_{p,0})))}\nonumber\\
 &\times\exp\Big(c_0
  \int_{x_0}^x
   \frac{\theta(\underline{z}(P_{\infty-},\underline{\hat{\mu}}(x^\prime,y,t_p)))}{\theta(\underline{z}(P_{\infty+},\underline{\hat{\mu}}(x^\prime,y,t_p)))}
   \frac{\theta(\underline{z}(P_{\infty+},\underline{\hat{\nu}}(x^\prime,y,t_p)))}{\theta(\underline{z}(P_{\infty-},\underline{\hat{\nu}}(x^\prime,y,t_p)))}dx^\prime\Big), \label{st5.42}\\
    r(x,y,t_{p})=&~
  -q(x_0,y_0,t_{p,0})\frac{\theta(\underline{z}(P_{\infty-}, \underline{\hat{\mu}}(x_0,y_0,t_{p,0})))}{\theta(\underline{z}(P_{\infty-}, \underline{\hat{\nu}}(x,y,t_{p})))}\frac{\theta(\underline{z}(P_{\infty+}, \underline{\hat{\nu}}(x,y,t_{p})))}{\theta(\underline{z}(P_{\infty+}, \underline{\hat{\mu}}(x_0,y_0,t_{p,0})))}\nonumber\\
  &~\times
  \exp\Big(-c_0\int_{x_0}^x
   \frac{\theta(\underline{z}(P_{\infty-},\underline{\hat{\mu}}(x^\prime,y,t_p)))}{\theta(\underline{z}(P_{\infty+},\underline{\hat{\mu}}(x^\prime,y,t_p)))}
   \frac{\theta(\underline{z}(P_{\infty+},\underline{\hat{\nu}}(x^\prime,y,t_p)))}{\theta(\underline{z}(P_{\infty-},\underline{\hat{\nu}}(x^\prime,y,t_p)))}dx^\prime\Big).\label{st5.43}\\
  q(x,y,t_p)&r(x,y,t_p)=-2c_0 \frac{\theta(\underline{z}(P_{\infty-},\underline{\hat{\mu}}(x,y,t_p)))}{\theta(\underline{z}(P_{\infty+},\underline{\hat{\mu}}(x,y,t_p)))}
   \frac{\theta(\underline{z}(P_{\infty+},\underline{\hat{\nu}}(x,y,t_p)))}{\theta(\underline{z}(P_{\infty-},\underline{\hat{\nu}}(x,y,t_p)))}.
   \end{align}
where $c_0=2e^{c_{\infty-}-c_{\infty+}}$ and
\begin{align}
  C_1(x,y,t_{p})
  =&~\frac{\theta(\underline{z}(P_{\infty+}, \underline{\hat{\mu}}(x_0,y_0,t_{p,0})))}{\theta(\underline{z}(P_{\infty+}, \underline{\hat{\mu}}(x,y,t_{p})))}\times\exp\Big( \frac{c_0}{2}\int_{x_0}^x
   \frac{\theta(\underline{z}(P_{\infty-},\underline{\hat{\mu}}(x^\prime,y,t_p)))}{\theta(\underline{z}(P_{\infty+},\underline{\hat{\mu}}(x^\prime,y,t_p)))} \nonumber\\
  &~\times
   \frac{\theta(\underline{z}(P_{\infty+},\underline{\hat{\nu}}(x^\prime,y,t_p)))}{\theta(\underline{z}(P_{\infty-},\underline{\hat{\nu}}(x^\prime,y,t_p)))}dx^\prime\Big), \\
    C_2(x,y,t_{p})
  =&~~~\frac{\theta(\underline{z}(P_{\infty-}, \underline{\hat{\mu}}(x_0,y_0,t_{p,0})))}{\theta(\underline{z}(P_{\infty-}, \underline{\hat{\mu}}(x,y,t_{p})))}
    \times\exp\Big(-\frac{c_0}{2}\int_{x_0}^x
   \frac{\theta(\underline{z}(P_{\infty-},\underline{\hat{\mu}}(x^\prime,y,t_p)))}{\theta(\underline{z}(P_{\infty+},\underline{\hat{\mu}}(x^\prime,y,t_p)))}\nonumber\\
  &~\times
   \frac{\theta(\underline{z}(P_{\infty+},\underline{\hat{\nu}}(x^\prime,y,t_p)))}{\theta(\underline{z}(P_{\infty-},\underline{\hat{\nu}}(x^\prime,y,t_p)))}dx^\prime\Big).
\end{align}
Here we emphasize that the initial values $q(x_0,y_0,t_{p,0})$ and $r(x_0,y_0,t_{p,0})$ should satisfy
\begin{align}\label{st5.47}
   r(x_0,y_0,t_{p,0})
=&~-q(x_0,y_0,t_{p,0})\frac{\theta(\underline{z}(P_{\infty-}, \underline{\hat{\mu}}(x_0,y_0,t_{p,0})))}{\theta(\underline{z}(P_{\infty-}, \underline{\hat{\nu}}(x_0,y_0,t_{p,0})))} \nonumber\\
&~\times\frac{\theta(\underline{z}(P_{\infty+}, \underline{\hat{\nu}}(x_0,y_0,t_{p,0})))}{\theta(\underline{z}(P_{\infty+}, \underline{\hat{\mu}}(x_0,y_0,t_{p,0})))}.
\end{align}
Moreover, the Abel map linearizes the auxiliary divisors
       $\mathcal{D}_{\hat{\underline{\mu}}(x,y,t_{p})},
       \mathcal{D}_{\hat{\underline{\nu}}(x,y,t_{p})}$
       in the sense that
         \begin{align}
         \underline{\alpha}_{Q_0}(\mathcal{D}_{\underline{\hat{\mu}}(x,y,t_{p})})
         =&~\underline{\alpha}_{Q_0}(\mathcal{D}_{\underline{\hat{\mu}}(x_0,y_0,t_{p,0})})
         +\underline{U}_0^{(2)}(x-x_0)+\underline{U}_{1}^{(2)}(y-y_{0})\nonumber\\
         &+ \underline{U}_{r}^{(2)}(t_{p}-t_{p,0}),\label{5.24m}\\
         \underline{\alpha}_{Q_0}(\mathcal{D}_{\underline{\hat{\nu}}(x,y,t_{p})})
         =&~\underline{\alpha}_{Q_0}(\mathcal{D}_{\underline{\hat{\mu}}(x_0,y_0,t_{p,0})})
         +\underline{U}_0^{(2)}(x-x_0)+\underline{U}_{1}^{(2)}(y-y_{0})\nonumber\\
         &+\underline{U}_{r}^{(2)}(t_{p}-t_{p,0})+\frac{1}{2}\Big(A_{Q_0}(P_{\infty-})-A_{Q_0}(P_{\infty+})\Big).\label{5.25m}
       \end{align}

\end{thm5.3}
\proof  First, we prove expression (\ref{5.15cp}). Denote the right hand of (\ref{5.15cp}) by $\tilde{\psi}_1$, and then one finds $\tilde{\psi}_1$ is meromorphic on $X\backslash\{P_{\infty\pm}\}$
with simple zeros at $\hat{\mu}_j(x,y,t_{p}),j=1,\ldots,n,$ and simple poles at $\hat{\mu}_j(x_0,y_0,t_{p,0}),j=1,\ldots,n,$ by Riemann vanishing theorem. Comparing (\ref{4.27dp}), (\ref{4.28dp}), (\ref{5.15cp}) for $\tilde{\psi}_1$, and taking into account (\ref{5.13}), (\ref{5.14}), one infers $\psi_1$ and $\tilde{\psi}_1$
have identical exponential behavior up to order $O(1)$ near $P_{\infty\pm}.$ Thus,
$\psi_1$ and $\tilde{\psi}_1$ share the same singularities and zeros and the Riemann-Roch-type
uniqueness result then proves that $\psi_1$ and $\tilde{\psi}_1$ coincide up to
normalization and we denote this normalization constant by $C_1(x,y,t_{p}).$ Hence $\psi_1(P,x,y,t_{p})$ has the form of (\ref{5.15cp}).
Using (\ref{4.1k}), (\ref{4.28dp}), (\ref{5.15cp}), one obtains
\begin{align}
  (\ln \psi_{1}(P,x,y,t_p))_x=&~-\eta+\frac{1}{4}q(x,y,t_p)r(x,y,t_p)\nonumber\\
  &~+q(x,y,t_p)\sqrt{\eta}\frac{\psi_2(P,x,y,t_p)}{\psi_1(P,x,y,t_p)}
\end{align}
and
\begin{align}
  -\frac{q(x,y,t_p)r(x,y,t_p)}{4}=&~\Big(\ln C_1(x,y,t_p)+\ln \frac{\theta(\underline{z}(P_{\infty+}, \underline{\hat{\mu}}(x,y,t_{p})))}{\theta(\underline{z}(P_{\infty+}, \underline{\hat{\mu}}(x_0,y_0,t_{p,0})))}
  \Big)_x \nonumber\\
  &~~~~~~~~~~~~~~~~~~~~~~~~~~~~~~~~~~~~~~~\textrm{as $P\rightarrow P_{\infty+}$,}\label{5.22m}\\
  \frac{q(x,y,t_p)r(x,y,t_p)}{4}=&~\Big(\ln C_2(x,y,t_p)+\ln \frac{\theta(\underline{z}(P_{\infty-}, \underline{\hat{\mu}}(x,y,t_{p})))}{\theta(\underline{z}(P_{\infty-}, \underline{\hat{\mu}}(x_0,y_0,t_{p,0})))}
  \Big)_x\nonumber\\
   &~~~~~~~~~~~~~~~~~~~~~~~~~~~~~~~~~~~~~~\textrm{as $P\rightarrow P_{\infty-}$.}\label{5.23m}
\end{align}
Then by Theorem 4.6 and Lemma 5.1, one infers
\begin{align}
   q(x,y,t_p)r(x,y,t_p)=&~
   -4e^{c_{\infty-}-c_{\infty+}}
   \frac{\theta(\underline{z}(P_{\infty-},\underline{\hat{\mu}}(x,y,t_p)))}{\theta(\underline{z}(P_{\infty+},\underline{\hat{\mu}}(x,y,t_p)))}
   \frac{\theta(\underline{z}(P_{\infty+},\underline{\hat{\nu}}(x,y,t_p)))}{\theta(\underline{z}(P_{\infty-},\underline{\hat{\nu}}(x,y,t_p)))}.
   \label{5.24m}
\end{align}
Thus, from (\ref{5.22m}), (\ref{5.23m}) and (\ref{5.16cp}), one gets
\begin{align}\label{5.21cc}
 C_1(x,y,t_{p})
  =&~\frac{\theta(\underline{z}(P_{\infty+}, \underline{\hat{\mu}}(x_0,y_0,t_{p,0})))}{\theta(\underline{z}(P_{\infty+}, \underline{\hat{\mu}}(x,y,t_{p})))}\times \exp\Big(-\frac{1}{4}\int_{x_0}^xq(x^\prime,y,t_p)r(x^\prime,y,t_p)dx^\prime\Big)\nonumber\\
  =&~\frac{\theta(\underline{z}(P_{\infty+}, \underline{\hat{\mu}}(x_0,y_0,t_{p,0})))}{\theta(\underline{z}(P_{\infty+}, \underline{\hat{\mu}}(x,y,t_{p})))}\times\exp\Big(e^{c_{\infty-}-c_{\infty+}} \nonumber\\
  &~\int_{x_0}^x
   \frac{\theta(\underline{z}(P_{\infty-},\underline{\hat{\mu}}(x^\prime,y,t_p)))}{\theta(\underline{z}(P_{\infty+},\underline{\hat{\mu}}(x^\prime,y,t_p)))}
   \frac{\theta(\underline{z}(P_{\infty+},\underline{\hat{\nu}}(x^\prime,y,t_p)))}{\theta(\underline{z}(P_{\infty-},\underline{\hat{\nu}}(x^\prime,y,t_p)))}dx^\prime\Big),\\
    C_2(x,y,t_{p})=&~\frac{\theta(\underline{z}(P_{\infty-}, \underline{\hat{\mu}}(x_0,y_0,t_{p,0})))}{\theta(\underline{z}(P_{\infty-}, \underline{\hat{\mu}}(x,y,t_{p})))}\times
     \exp\Big(\frac{1}{4}\int_{x_0}^xq(x^\prime,y,t_p)r(x^\prime,y,t_p)dx^\prime\Big)\nonumber\\
    =&~~\frac{\theta(\underline{z}(P_{\infty-}, \underline{\hat{\mu}}(x_0,y_0,t_{p,0})))}{\theta(\underline{z}(P_{\infty-}, \underline{\hat{\mu}}(x,y,t_{p})))}
    \times\exp\Big(-e^{c_{\infty-}-c_{\infty+}} \nonumber\\
  &~\int_{x_0}^x
   \frac{\theta(\underline{z}(P_{\infty-},\underline{\hat{\mu}}(x^\prime,y,t_p)))}{\theta(\underline{z}(P_{\infty+},\underline{\hat{\mu}}(x^\prime,y,t_p)))}
   \frac{\theta(\underline{z}(P_{\infty+},\underline{\hat{\nu}}(x^\prime,y,t_p)))}{\theta(\underline{z}(P_{\infty-},\underline{\hat{\nu}}(x^\prime,y,t_p)))}dx^\prime\Big).
 \end{align}
On the other hand, taking into account the asymptotic behavior of both sides in  (\ref{4.19kp}) and (\ref{4.20kp}) for $P\rightarrow P_{\infty+}$, one infers
\begin{align}
   q(x,y,t_{p})=&~q(x_0,y_0,t_{p,0}) C_1(x,y,t_{p})^2\frac{\theta(\underline{z}(P_{\infty-}, \underline{\hat{\mu}}(x,y,t_{p})))}{\theta(\underline{z}(P_{\infty-}, \underline{\hat{\mu}}(x_0,y_0,t_{p,0})))}\nonumber\\
   &\times
  \frac{\theta(\underline{z}(P_{\infty+}, \underline{\hat{\mu}}(x,y,t_{p})))}{\theta(\underline{z}(P_{\infty+}, \underline{\hat{\mu}}(x_0,y_0,t_{p,0})))} \nonumber\\
 =&~q(x_0,y_0,t_{p,0}) \frac{\theta(\underline{z}(P_{\infty+}, \underline{\hat{\mu}}(x_0,y_0,t_{p,0})))}{\theta(\underline{z}(P_{\infty+}, \underline{\hat{\mu}}(x,y,t_{p})))} \nonumber\\
 &\times
  \frac{\theta(\underline{z}(P_{\infty-}, \underline{\hat{\mu}}(x,y,t_{p})))}{\theta(\underline{z}(P_{\infty-}, \underline{\hat{\mu}}(x_0,y_0,t_{p,0})))}\times\exp\Big(2e^{c_{\infty-}-c_{\infty+}} \nonumber\\
  &~\int_{x_0}^x
   \frac{\theta(\underline{z}(P_{\infty-},\underline{\hat{\mu}}(x^\prime,y,t_p)))}{\theta(\underline{z}(P_{\infty+},\underline{\hat{\mu}}(x^\prime,y,t_p)))}
   \frac{\theta(\underline{z}(P_{\infty+},\underline{\hat{\nu}}(x^\prime,y,t_p)))}{\theta(\underline{z}(P_{\infty-},\underline{\hat{\nu}}(x^\prime,y,t_p)))}dx^\prime\Big), \label{5.22abc}
   \end{align}
   and
   \begin{align}
    r(x,y,t_{p})=&~-q(x_0,y_0,t_{p,0}) C_2(x,y,t_{p})^2\frac{\theta(\underline{z}(P_{\infty-}, \underline{\hat{\nu}}(x,y,t_{p})))}{\theta(\underline{z}(P_{\infty-}, \underline{\hat{\mu}}(x_0,y_0,t_{p,0})))}\nonumber\\
   & \times
  \frac{\theta(\underline{z}(P_{\infty+}, \underline{\hat{\nu}}(x,y,t_{p})))}{\theta(\underline{z}(P_{\infty+}, \underline{\hat{\mu}}(x_0,y_0,t_{p,0})))} \nonumber\\
   =&~-q(x_0,y_0,t_{p,0})\frac{\theta(\underline{z}(P_{\infty-}, \underline{\hat{\mu}}(x_0,y_0,t_{p,0})))}{\theta(\underline{z}(P_{\infty-}, \underline{\hat{\nu}}(x,y,t_{p})))}\frac{\theta(\underline{z}(P_{\infty+}, \underline{\hat{\nu}}(x,y,t_{p})))}{\theta(\underline{z}(P_{\infty+}, \underline{\hat{\mu}}(x_0,y_0,t_{p,0})))}\nonumber\\
  &~\times
  \exp\Big(-2e^{c_{\infty-}-c_{\infty+}}\int_{x_0}^x
   \frac{\theta(\underline{z}(P_{\infty-},\underline{\hat{\mu}}(x^\prime,y,t_p)))}{\theta(\underline{z}(P_{\infty+},\underline{\hat{\mu}}(x^\prime,y,t_p)))}
   \frac{\theta(\underline{z}(P_{\infty+},\underline{\hat{\nu}}(x^\prime,y,t_p)))}{\theta(\underline{z}(P_{\infty-},\underline{\hat{\nu}}(x^\prime,y,t_p)))}dx^\prime\Big).\label{5.22abcd}
  \end{align}
  Moreover, (\ref{st5.47}) follows from (\ref{5.22abc}) and (\ref{5.22abcd})
  by taking $(x,y,t_{p})=(x_0,y_0,t_{p,0}).$
 Finally,
the linearization property of the
Abel map in (\ref{5.24m}) and (\ref{5.25m}) is a standard investigation of the differentials $$\Omega_i(x,y,t_{p})=
d \ln(\psi_i(\cdot,x,y,t_{p})),~i=1,2,$$
or standard Langrange interpolation procedure (see \cite{Gesztesyhoden2003},~\cite{Novikov}). \qed \vspace{0.3cm}

Thus, the results of theorem 2.1 and theorem 5.2 enable us to
construct the algebro-geometric solutions of the whole mKP hierarchy.
\newtheorem{thm5.4}[thm5.1]{Theorem}
\begin{thm5.4}
Assume spectral curve defined in (\ref{2.22}) is nonsingular and $q,r$ satisfy (\ref{st2.51}).
Moreover, let $P\in X\backslash\{P_{\infty\pm}\}$ and the divisors $\mathcal{D}_{\underline{\hat{\nu}}(x,y,t_{p})}$, $\mathcal{D}_{\underline{\hat{\mu}}(x,y,t_{p})}$ are nonspecial. The algbro-geometric solutions of $p$th mKP equation ($p\geq 2$)
possess the following theta function representation

     \begin{equation}\label{solution1}
  u(x,y,t_p)=2^pc_0 \frac{\theta(\underline{z}(P_{\infty-},\underline{\hat{\mu}}(2x,-y,-(-2)^{p-1}t_p)))}{\theta(\underline{z}(P_{\infty+},\underline{\hat{\mu}}(2x,-y,-(-2)^pt_p)))}
   \frac{\theta(\underline{z}(P_{\infty+},\underline{\hat{\nu}}(2x,-y,(-2)^{p-1}t_p)))}{\theta(\underline{z}(P_{\infty-},\underline{\hat{\nu}}(2x,-y,-(-2)^pt_p)))}.
\end{equation}
where the constant $c_0\in\mathbb{C}$ is the same with that in theorem 5.2.

\end{thm5.4}

\newtheorem{fin2}[thm5.1]{Remark}
\begin{fin2}

Here we consider the special case $p=2.$\\
(i)
In the case $p=2,$ the expression (\ref{solution1})
yields algebro-geometric solutions of the mKP equation and
 real algebro-geometric solutions of the mKP equation can easily obtained from (\ref{st5.42}) and (\ref{st5.43})
 by studying reduction condition $q=\pm\bar{r}$.
  \\
(ii) The spectral curve of the mKP equation (\ref{1.1kp}) is hyperelliptic which compactified by two different points $P_{\infty\pm}$
at infinity. Similarly we can discuss algebro-geometric
solution of the mKP hierarchy on some
specific trigonal curves following this way.
Moreover, if the solution (\ref{solution1}) is independent of $y$,
then we can derive theta function representations for algebro-geometric solutions
of modified KdV equation.

 \end{fin2}


\section*{Acknowledgements}
The work described in this paper was supported by grants from the National
Science Foundation of China (Project No. 10971031), and the Shanghai
Shuguang Tracking Project (Project No. 08GG01).

\end{document}